\documentclass{cernyrep} 
\usepackage{texnames}
\usepackage[T1]{fontenc}
\usepackage{hyperref}
\hypersetup{
    colorlinks=true,
    citecolor=black,
    linkcolor=black,
    filecolor=black,      
    urlcolor=blue,
}
\usepackage{fancyhdr}
\fancyhfoffset{4mm}
\fancypagestyle{ARTTITLE}{%
\fancyhf{} 
\lhead{\small{Proceedings of the 2018 CERN--Accelerator--School course on \it{Beam Instrumentation}, Tuusula, (Finland)}} 
\lfoot{Available online at \url{https://cas.web.cern.ch/previous-schools}}
\rfoot{\thepage\hspace*{3mm}}

}

\begin{document}
\title{Transverse beam profiles}
 
\author{E. Bravin}

\institute{CERN, Geneva, Switzerland}

\begin{abstract}
The performance and safe operation of a particle accelerator is closely
connected to the transverse emittance of the beams it produces. For
this reason many techniques have been developed over the years for
monitoring the transverse distribution of particles along accelerator
chains or over machine cycles. The definition of beam profiles is explained and
the different techniques available for the detection
of the particle distributions are explored. Examples of concrete applications of
these techniques are given.
\end{abstract}
\maketitle

\thispagestyle{ARTTITLE} 
 
\section{Introduction}
 
Before discussing the details of beam profile measurements it is essential to
understand what we mean by beam profiles. In the following sections
the concepts of \emph{transverse
planes} and \emph{beam profiles} will therefore be
explained.

\subsection{Coordinate system}
First of all we need to define the coordinate system. What we usually call
\emph{beams} or \emph{bunches} are in fact a collection
of a very large number of particles whose centre of gravity moves in a
well defined direction. In the ideal machine the average direction of
motion of the particles is called the ideal orbit and in general passes through the
centre of the focusing elements of the machine.

\begin{figure}[htb]
\centering\includegraphics[width=0.3\linewidth]{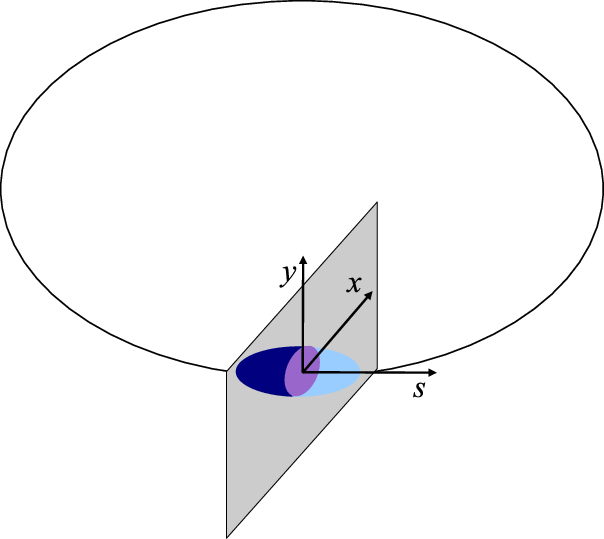}
\caption{Reference coordinate system}
\label{fig:referencesystem}
\end{figure}

\Figure[b]~\ref{fig:referencesystem} shows a typical reference frame used in
accelerators. The average direction of motion is used to define the
longitudinal axis and is often called $s$. Once the $s$
axis is defined we can define a plane orthogonal to it, this is the
transverse plane. On this transverse plane we define the other two
axes, one orthogonal to the plane containing the ideal orbit, this is the
$y$ axis or vertical axis and usually points upward, and one,
the $x$ or horizontal axis, orthogonal to the other two and forming with them a right-handed reference system.

\subsection{Transverse geometrical space}
In the previous section the concept of transverse plane was
introduced, in this section we shall introduce the concept of the
geometrical transverse space or $x$, $y$ space.

\begin{figure}[htb]
\centering\includegraphics[width=0.5\linewidth]{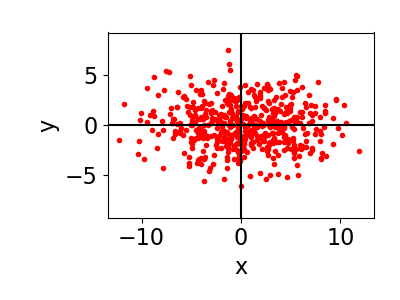}
\caption{Particle distribution in the transverse $x$, $y$ space}
\label{fig:distributioninxyspace}
\end{figure}

Consider a transverse plane, every time a particle crosses this plane we take note of the
$x$ and $y$ coordinates of the particle. We then transfer
these points onto a 2D chart and obtain the particle distribution in
the $x$, $y$ space as shown in \Fref{fig:distributioninxyspace}.

If there is no coupling between the motion of the particles in the $x$ and in the $y$ direction this distribution will be somehow elliptical with the axes of the ellipse parallel to the $x$ and $y$ axes.

\subsection{Transverse phase space}
\begin{figure}[htb]
\centering\includegraphics[width=0.4\linewidth]{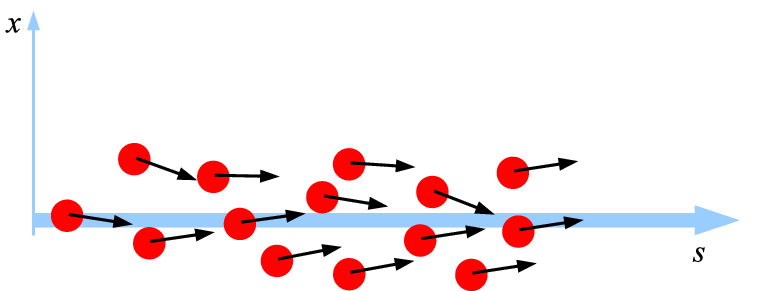}
\centering\includegraphics[width=0.4\linewidth]{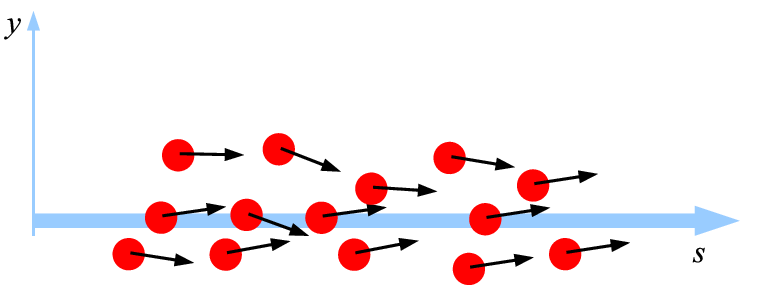}
\caption{Each particle in a beam moves in a slightly different direction}
\label{fig:particlesrandommovement}
\end{figure}

As already mentioned before, a beam consists of many particles
whose centre of gravity moves in a well defined direction describing
the ideal orbit. In fact each particle of the beam moves in a slightly
different direction, see \Fref{fig:particlesrandommovement}, and focusing elements are necessary to keep the
particles close together.

\begin{figure}[htb]
\centering\includegraphics[width=0.6\linewidth]{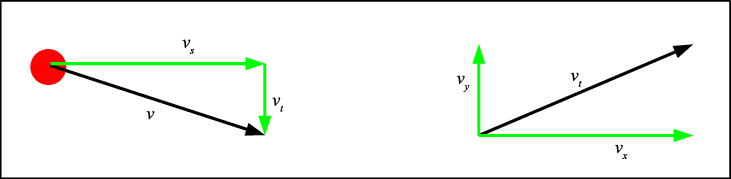}
\caption{Decomposition of the velocity vectors into longitudinal and transverse components}
\label{fig:velocitycomponents}
\end{figure}

The velocity vector of each particle can be decomposed into two
components: one parallel to the beam direction
$s$, and one orthogonal to it, the transverse velocity. The
transverse velocity can in its turn be decomposed into two more
components: one along $x$ and one along $y$ as shown in \Fref{fig:velocitycomponents}.

As in the previous exercise we can assume a transverse plane and
this time when a particle crosses it we note not only the $x$
and $y$ positions, but also the transverse component of the
momentum, $p_x$ and
$p_y$.

\begin{figure}[htb]
\centering\includegraphics[width=0.4\linewidth]{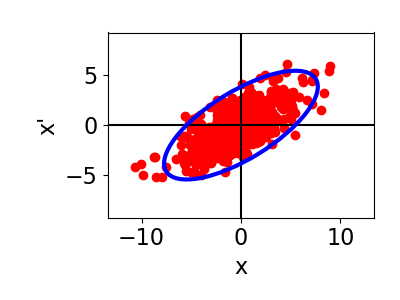}
\centering\includegraphics[width=0.4\linewidth]{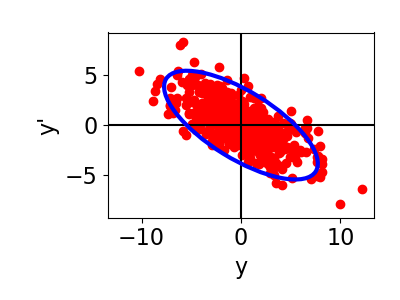}
\caption{Particle distribution in the transverse phase spaces with the 95\% confidence ellipses}
\label{fig:transverspacedistribution}
\end{figure}

In the previous exercise we used one 2D chart to plot the points, this time we need two charts, on one we plot for each particle a point corresponding to the $x$ position and the
$p_x$ momentum component, and on the second we do the same for the $y$ components. For convenience we normalise the transverse momenta by the longitudinal momentum $p_s$ with $x^\prime= p_x/p_s$ and $y^{\prime}= p_y/p_s$.
\Figure[b]~\ref{fig:transverspacedistribution} shows a sketch of typical transverse phase space distributions together with the smallest ellipses that contain 95\% of the particles.

The particle distributions in the phase space are again of elliptical
shape, but this time the ellipses can be tilted. Moreover the area of
the surface that contains all the particles is an invariant, this means
that if the same exercise is repeated at several locations around the
accelerator the ellipses will have different aspect ratios and
different tilts, but the area will remain the same.

To avoid confusion it is worth clarifying that in a machine where there
is no coupling between the $x$ and $y$ axes (ideal
machine) the two phase space distributions are completely independent,
and although the areas are invariant, the value is not necessarily the
same for the $x$ phase space and the $y$ phase space.

Because these two areas are invariants of motion they constitute
fundamental parameters that define the quality of the beam. These areas
are called emittance and more precisely horizontal emittance, indicated with $\varepsilon_x$, and vertical emittance,
indicated with $\varepsilon_y$.
In reality there are several definitions of the emittance, but in
practice all are related to the area of these ellipses.

It is worth mentioning that there are processes that can modify the transverse
emittances of a beam, like acceleration, blow-up and cooling, but these
topics are outside the scope of this lecture.

\subsection{Differences between $x$, $y$ space and \emph{phase space}}
As we have seen the $x$, $y$ space and the phase space
are different things and sometimes confusion arises between the two
when one first starts studying this subject.

The $x$, $y$
space only contains the spatial information of the distribution of the
particles in a beam and merges both $x$ and $y$ axis in a
single chart, while the two phase spaces $x$, $x^\prime$ and $y$, $y^\prime$
contain both the spatial and the velocity information.

The phase space contains the whole description of the state of all the
particles for that particular plane (initial conditions) and this is all that is required for calculating the
subsequent motion of the particle in the electromagnetic fields of the
accelerator.

Despite being different things, there are also common points between the
two spaces. In fact both contain the position information and thus they
have common projections. If one projects any of these 2D distributions
along the $x$ or $y$ axis one obtains exactly the same
thing. So the projection of the $x$, $y$ space along $x$ is
the same thing as the projection of the $x$, $x^\prime$ space along
$x$ and the same can be said for $y$.

These common properties allow us to sample the beam distribution in one
space and convert it to the other by using beam dynamics theories. This
is of great importance as sampling the $x$, $y$ space is
much easier than sampling the phase space, although what one really
needs is the phase space.

In practice the sampling of the phase space in a direct way is not
technically feasible and it is always approximated by measuring a
sequence of position distributions. Many different techniques,
described in other lectures in this school, have been developed over
the years to do this.

\subsection{Beam profiles}
Transverse beam profiles are simply histograms expressing the number of
particles in a beam as a function of the transverse position, thus we
have a horizontal profile expressing the number of particles at
different $x$ positions and we have a vertical profile
expressing the number of particles at different $y$ positions.

Theoretically these histograms could be considered as continuous
functions expressing particle densities vs. position. In reality, given
the discrete nature of particles and the discrete position binning of
the sampling, any profile measurement consists of a 1D histogram.
Considering the 2D distributions introduced early on, the profiles are
simply the projections of these 2D plots along one axis.

\begin{figure}[htb]
\centering\includegraphics[width=0.9\linewidth]{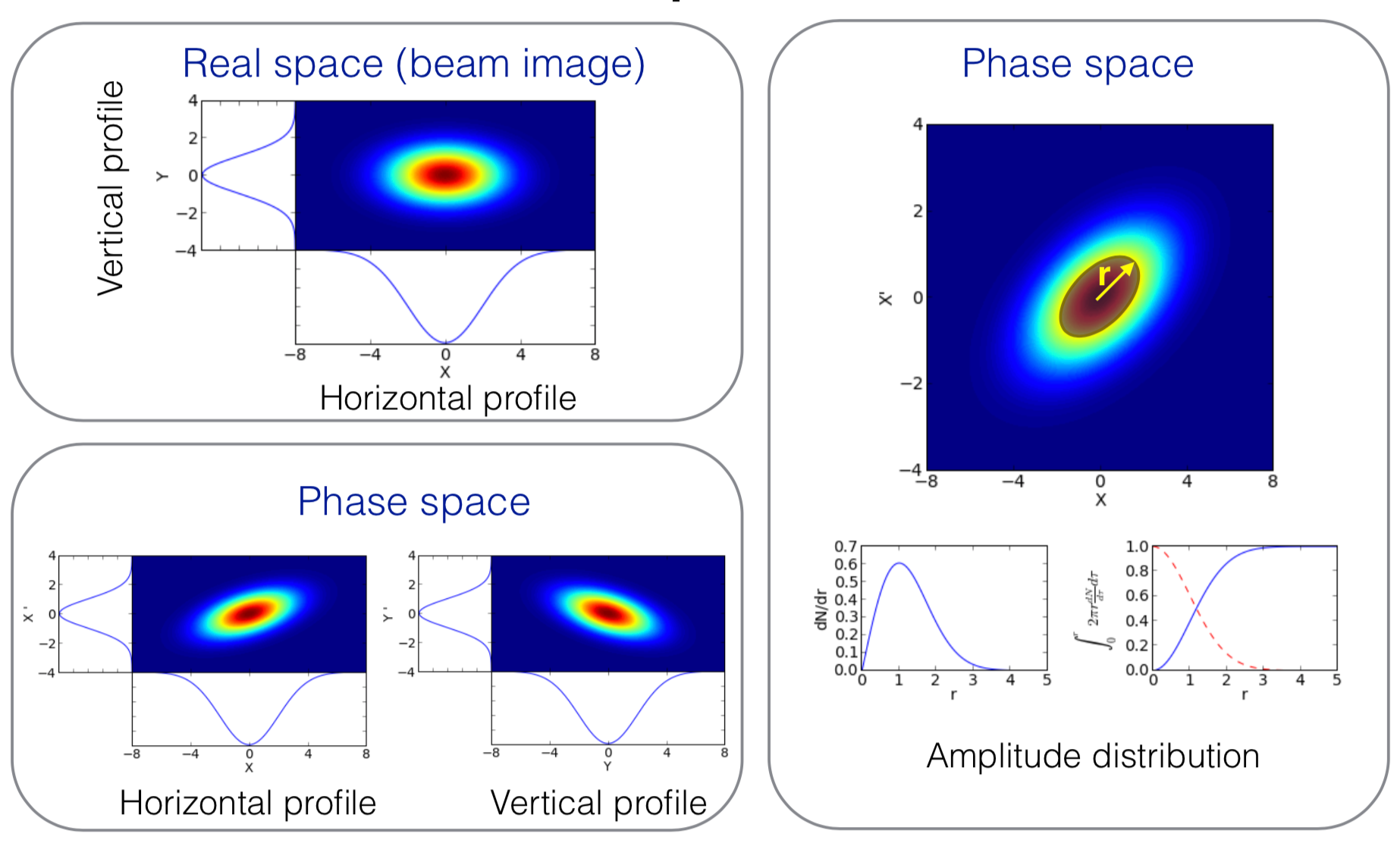}
\caption{Different types of 2D distributions and relative transverse profiles:
 $x$, $y$ transverse space (top left), transverse phase spaces (bottom left),
transverse phase space with amplitude distribution profile (right).}
\label{fig:twoddistributions}
\end{figure}

In the continuous case we can define a 2--dimensional density function
expressing the transverse distribution of the particles in the beam
$i(x,y)$ and the profiles are simply given by

\begin{align}
\text{Profile}_\text{H} (x) & =\int_{-\infty }^{+\infty} i(x,y) dy \label{eq:profxdisc}, \\
\text{Profile}_\text{V} (y) & =\int_{-\infty }^{+\infty } i(x,y) dx \label{eq:profydisc}.
\end{align}

In a real case the integration limits can of course be reduced to the size of the beam pipe. Typically the particles are distributed according to the Gaussian function

\begin{equation}
i(x,y) =\frac{N_0}{2 \pi \sigma_x\sigma_y} e^{- \left( \frac{x^2}{2 \sigma_x^2}+\frac{y^2}{2 \sigma_y^2} \right) },
\label{eq:bigaussian}
\end{equation}

and thus we have

\begin{align}
\text{Profile}_\text{H} (x) & =\frac{N_{0}}{\sqrt{2\pi }\sigma_{x}} e^{-{\frac{x^{2}}{2\sigma_{x}^{2}}}} \label{eq:profxcont}, \\
\text{Profile}_\text{V} (y) & =\frac{N_{0}}{\sqrt{2\pi }\sigma_{y}} e^{-{\frac{y^{2}}{2\sigma_{y}^{2}}}} \label{eq:profycont}.
\end{align}

\subsection{Normalised phase space and amplitude distribution}

There is also another type of transverse
profile that is sometimes used: it is called the amplitude distribution
and it represents a distribution in the phase space. At a given location in a circular
machine, turn after turn, a particle will evolve around a well defined ellipse in
phase space. Different particles will describe different ellipses, all similar differing only by a scale factor the \emph{amplitude}. 
It is thus possible to define the density of particles as a function
of this amplitude (in case the ellipse degenerates to a circle the amplitude is nothing else than the radius).

This sampling of the amplitude density is usually performed using scrapers and requires recording
either the beam losses or the beam current decrease as
function of the scraper position while moving the scrapers into the beam. \Figure[b]~\ref{fig:twoddistributions} shows a
sketch of the different 2D distributions and the relative profiles.

The distribution of the particles in phase space can generally be described by a tilted, by-dimensional, Gaussian function. For the calculation of the amplitude distribution it is convenient to apply simple transformations. We have already seen that for each particle we can define an ellipse in the phase space, as the particle moves along the accelerator this ellipse will evolve, but the area (emittance) will remain constant. By amplitude distribution we really mean the density of particles as function of their emittance ellipse and, by the invariant property of the emittance, it does not depend on the longitudinal position where we sample it. 
We can take advantage of this and apply a \emph{rotation} of the phase space so that the major and minor axis of the ellipse become parallel to the new $x$, $x^\prime$ axes, this is accomplished by applying the transport matrix to an opportune location where we have a waist (can be a virtual location).
Second we apply a scaling of the axes so that the ellipse becomes a circle, this is equivalent to applying a change of units (we can call these normalised units).
In this new phase space it is easy to define the radial particle distribution and compute the amplitude distribution

\begin{align}
    &i(x,x^\prime)= \frac{N_0}{2 \pi \sigma_x \sigma_{x^\prime} \sqrt{1 - \rho^2}}
    e^{-\frac{1}{2(1-\rho^2)} \left( \frac{x^2}{\sigma_x^2} +
    \frac{{x^\prime}^2}{\sigma_{x^\prime}^2} -
    \frac{2 \rho x x^\prime}{\sigma_x \sigma_{x^\prime}}\right)}
    \label{eq:phasespacecont},\\
    &i_\text{waist}(x,x^\prime)= \frac{N_0}{2 \pi \sigma_x \sigma_{x^\prime}}
    e^{-\frac{1}{2} \left( \frac{x^2}{\sigma_x^2} + \frac{{x^{\prime}}^2}{\sigma_{x^\prime}^2}\right)}
    \label{eq:phasespacewaitscont},\\
    &i_\text{waist normalised}(x_n,x_n^\prime)= \frac{N_0}{2 \pi \sigma_n^2}
    e^{- \frac{x_n^2 + {x_n^\prime}^2}{ 2\sigma_n^2}}
    \label{eq:phasespacenormcont},\\
    &r^2= x_n^2 + {x_n^\prime}^2 \label{eq:normalisedradius},\\
    &i_\text{waist normalised}(r)= \frac{N_0}{2 \pi \sigma_n^2}
    e^{- \frac{r^2}{ 2\sigma_n^2}} \label{eq:phasespacenormradiuscont},
\end{align}

where $\rho$ is the correlation coefficient between $x$ and $x^\prime$.
We can now calculate the density of particles as function of the amplitude $dN(r)/dr$ and the amplitude distribution $I(r)$

\begin{align}
    &\frac{dN(r)}{dr}= 2 \pi i(r) r = \frac{N_0}{\sigma^2} r
    e^{-\frac{r^2}{ 2\sigma^2}}
    \label{eq:amplitudedensityfunction},\\
    &I(r)= \int_0^r \frac{N_0}{\sigma^2} \tau
    e^{-\frac{\tau^2}{2\sigma^2}} d\tau= N_0\left(1-e^{-\frac{r^2}{ 2\sigma^2}}\right)
    \label{eq:amplitudefunction}.
\end{align}

\section{Sampling of distributions}
\subsection{Intercepting vs. non--intercepting}
The methods used to sample the distributions of particles in a beam can
be divided into two main categories: intercepting and non--intercepting.
As the names suggest, in the first group we have the methods that
require placing an object on the beam path and analysing the signal
produced by the interaction of the particles with the object, in the
second group we have the methods that use signals emitted spontaneously
by the beam. An exception to this definition is the laser wire scanner which is classified
as non--intercepting even though it requires an external stimulus (laser
beam), nevertheless, given the extremely small perturbation that this
stimulus has on the beam, it is neglected.
Among the intercepting techniques there are:

\begin{itemize}
\item scanning wires
\item wire grids
\item radiative screens
\end{itemize}
and among the non--intercepting:

\begin{itemize}
\item synchrotron light
\item rest gas ionisation
\item (inverse) Compton scattering
\item photo dissociation
\end{itemize}

\subsection{Interaction of particles with matter}
The most used techniques for sampling profiles are based on
intercepting methods; the understanding of the interactions between the
particles and the obstacles used in the sampling is essential.

There are many physical effects that can be exploited for the detection:

\begin{itemize}
\item ionisation
\begin{itemize}
\item creation of electron--ion pairs
\item secondary electron emission (SEM, low energy)
\item emission of photons
\end{itemize}

\item elastic and inelastic scattering
\begin{itemize}
\item dislocations
\item production of secondary particles (high energy)
\end{itemize}

\item \v{C}erenkov radiation
\item bremsstrahlung
\item optical transition radiation (OTR)
\end{itemize}

\subsubsection{Energy deposition}
The study of the energy deposited by the particles in our
\emph{sensor} (the object or obstacle we introduce on the beam path) has two important uses:

\begin{itemize}
\item Signals are often proportional to the amount of energy deposited
in the material (or more often the density of the energy deposition).
\item The deposited energy can be dangerous for the
\emph{sensor} and can destroy it, again the
total energy and the energy density are both important.
\end{itemize}

The quantity widely used to describe the energy deposition is the
$dE/dx$. This quantity describes the amount of energy lost by a
particle in traversing a unit length of material, often this quantity
is normalised for the material density and the units become thus
$\UMeV \Unit{}{cm}^2 / \Unit{}{g}$.

The mechanism of energy deposition by charged particles in matter in
the range in which we are interested is well described by the Bethe--Bloch
formula which gives the energy deposition density as a function of
particle charge, particle momentum and material properties:

\begin{equation}
-\frac{dE}{dx}= K z^2 \frac{Z}{A}\frac{1}{\beta^2} \left[ \frac{1}{2} ln \frac{2 m_e c^2 \beta^2 \gamma^2 T_{max}}{I^2} - \beta^2 - \frac{\delta(\beta\gamma)}{2} \right].
\label{ eq:bethebloch}
\end{equation}

\begin{figure}[htb]
\centering\includegraphics[width=0.8\linewidth]{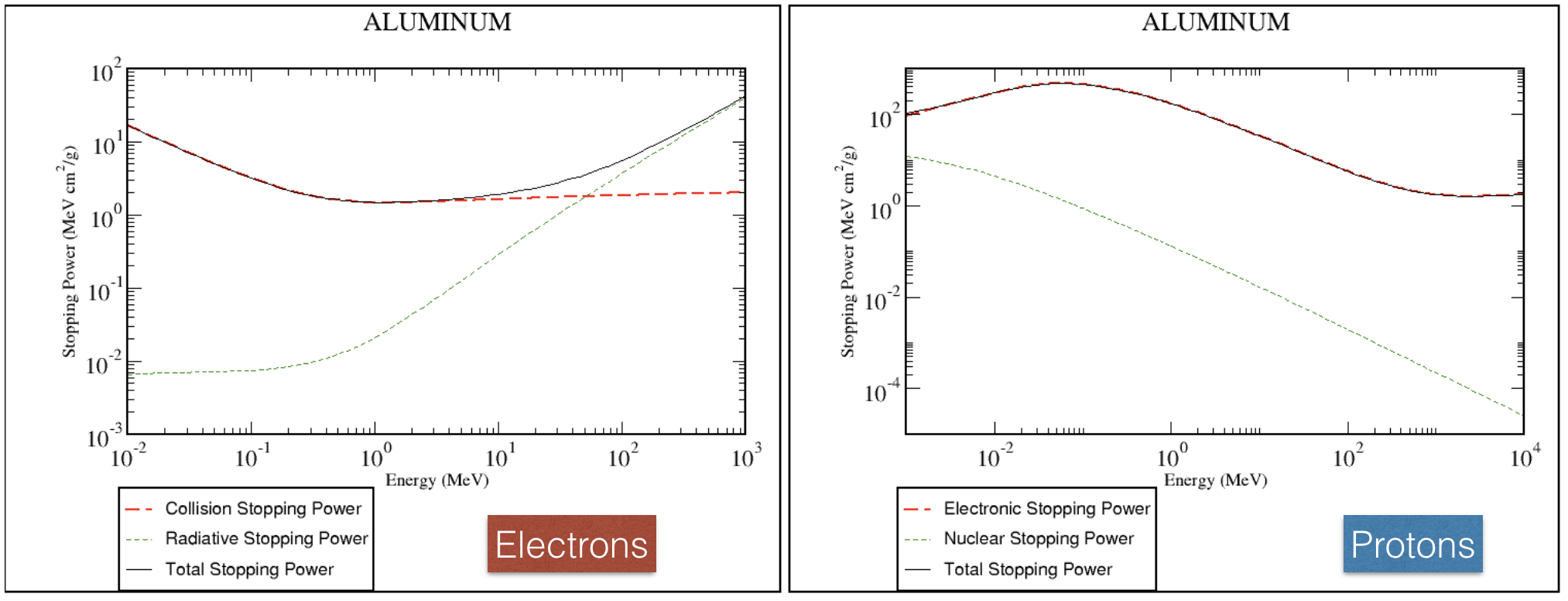}
\caption{The $dE/dx$ curve for electrons and protons in aluminium. Data from https://physics.nist.gov}
\label{fig:dedxaluminium}
\end{figure}

It should be mention that the Bethe--Bloch formula is not valid for electrons. Most of the interactions between the particles and the material happens in fact with the electrons inside the material so that in case the particle is an electron it is not possible to distinguish between the target and the projectile. For electrons other models or tabulated values should be used. The overall shape of the $dE/dx$ function is however very similar to the Bethe--Bloch formula in the energy range we usually operate. \Figure[b]~\ref{fig:dedxaluminium} shows the energy deposition curves for protons and electrons in aluminium, note that the energy scales are not the same.

Moreover the Bethe--Bloch formula
does not describe the amount of energy deposited in the material, but the
amount of energy lost by the primary particle. This has an
important consequence for our purposes as often our
\emph{sensors} are very thin and
absorb only a small fraction of the energy lost by the particle in
traversing it. Think, for example, of the case of thin metallic foils
traversed by high--energy electrons, many X-rays will be produced but
only a negligible fraction of the energy of the X-rays will be
reabsorbed in the foil.

\begin{figure}[htb]
\centering\includegraphics[width=0.8\linewidth]{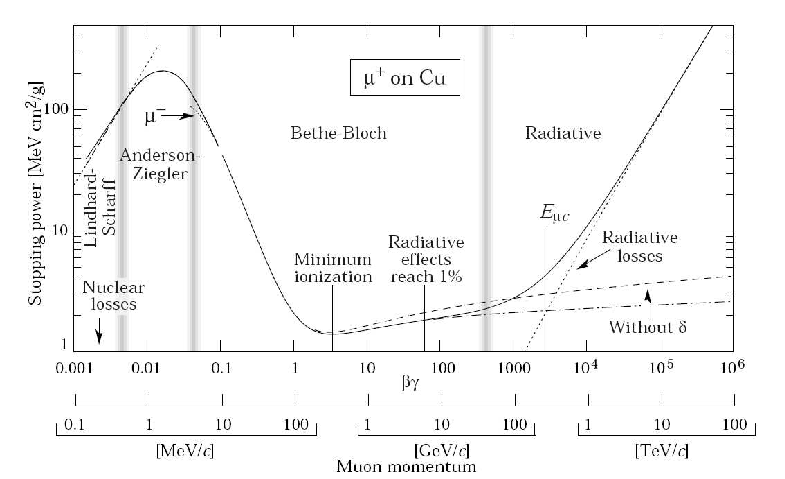}
\caption{The $dE/dx$ curve for positive muons in copper, the same curve describes very
well other elementary particles with $q= 1$ and different materials.
Note that the horizontal axis is given in terms of particle momentum
and not particle energy. For reference, $\beta\gamma= 1$
corresponds to a kinetic energy of $212\UkeV$ for electrons and $390\UMeV$
for protons.}
\label{fig:betheblock}
\end{figure}

The $dE/dx$ curve (an example is shown in 
\Fref{fig:betheblock}) has been found to be almost independent of the
material properties and the particle type (for $q=\pm1$ elementary
particles) provided one uses the particle momentum and not the particle
energy (it is the speed of the particle that matters in the ionisation
process). The only parameter of the material important for the energy
deposition is the density and it disappears if the $dE/dx$ is normalised for
it.

An important aspect to notice is that at low momentum the
$dE/dx$ is much larger, this means that for low--energy beams we
can expect larger signals, but also more important thermal problems.

Looking at the $dE/dx$ curve we notice that the $dE/dx$ increases rapidly at very low particle energies,
reaches a maximum for $\beta\gamma \sim 0.02$,
then decreases rapidly and reaches a minimum for
$\beta\gamma \sim 4$ (minimum ionising particle MIP); from there on it
increases only slowly (not considering the radiative part). As a
reference, $\beta\gamma=0.02$ corresponds to
protons of $188\UkeV$ and electrons of $100\UeV$ kinetic
energy and $\beta\gamma=4$ corresponds to
protons of $2.9\UGeV$ and electrons of $1.6\UMeV$
kinetic energy. From these values it is clear that in the field of
instrumentation for particle accelerators the relevant part of the
$dE/dx$ curve is the one described by the Bethe--Bloch (again excluding
radiative effects) that is valid for
$\beta\gamma>0.05$. 

The main process occurring when a charged particle traverses a
material is the creation of electron--ion pairs and the creation of
excited states in atoms. The electron--ion creation is
often used to detect the passage of the particles, this is the case for
example of ionisation chambers and solid--state detectors. In these
cases a bias voltage is applied across the material in order to attract
the positive and negative charges on opposite electrodes and measure
the resulting current.

\subsubsection{Secondary Emission (SEM)}
In the ionisation process ion--electron pairs are produced all along the
particle track, with a wide distribution of kinetic energy
being transferred to the liberated electrons. In some cases the pairs are
produced near the surface of the material and the electron has
sufficient momentum and the right direction to reach the surface. If
the energy of the electron is larger than the surface potential barrier it may escape in the surrounding vacuum as depicted in \Fref{fig:semandscintillation}--a. 
This phenomenon can be amplified by biasing the material with a negative
voltage, this will reduce the potential barrier at the surface and also
prevent free electrons from being recaptured by the material.

Only a very thin layer near the surface of the material is involved in
secondary emission, this means that it is not the bulk
properties of the material that matter but the properties of the surface layer,
commonly an oxide form of the bulk material that can be modified by the irradiation (ageing).

\begin{figure}[htb]
\centering\includegraphics[width=0.6\linewidth]{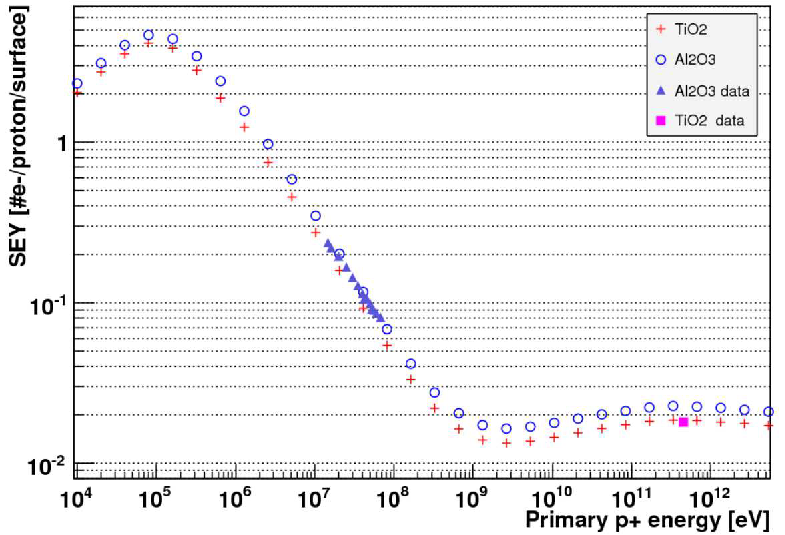}
\caption{Calculated and measured secondary emission yield for protons in
different materials as function of proton kinetic energy.}
\label{fig:semmeasurement}
\end{figure}

Usually SEM yields are measured experimentally for the different
materials and different incident energies. \Figure[b]~\ref{fig:semmeasurement}
shows such a measurement for protons on titanium and aluminium oxides
together with the analytical predictions based on the model described
in Ref. \cite{PhysRev.107.977}. As you can see, the measured points fit very well with the
prediction and nicely follow the $dE/dx$ curve.

\subsubsection{Scintillation}
\begin{figure}[htb]
\centering
\begin{tabular}{c @{\hspace{1cm}} c}
\includegraphics[width=0.2\linewidth]{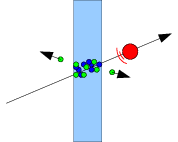} &
\includegraphics[width=0.2\linewidth]{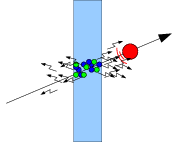} \\
a) & b) \\
\end{tabular}
\caption{A charged particle crosses a material and leaves behind a track of
ion--electron pairs. a) Free electrons created near the surface can have
sufficient energy to reach the surface and escape the material. b) Excited atoms and molecules decay to lower levels emitting photons.}
\label{fig:semandscintillation}
\end{figure}

The ion--electron pairs produced by the passage of a charged particle
will eventually recombine. When this recombination happens, the binding
energy of the electron will be emitted in the form of electromagnetic
radiation (photons). The passing primary particle will also excite
atoms and molecules without ionising them, in this case, again, the
return to the ground state is accompanied by the emission of photons, see \Fref{fig:semandscintillation}--b.
The de-excitation process can involve many meta-stable levels and only a
few of the transitions will emit photons in a range suitable for
observation, typically the visible range. The lifetime of the excited
states can vary over a large range and the emission can thus span over
relatively long periods and will depend on the properties of the
material.

\begin{figure}[htb]
\centering\includegraphics[width=0.6\linewidth]{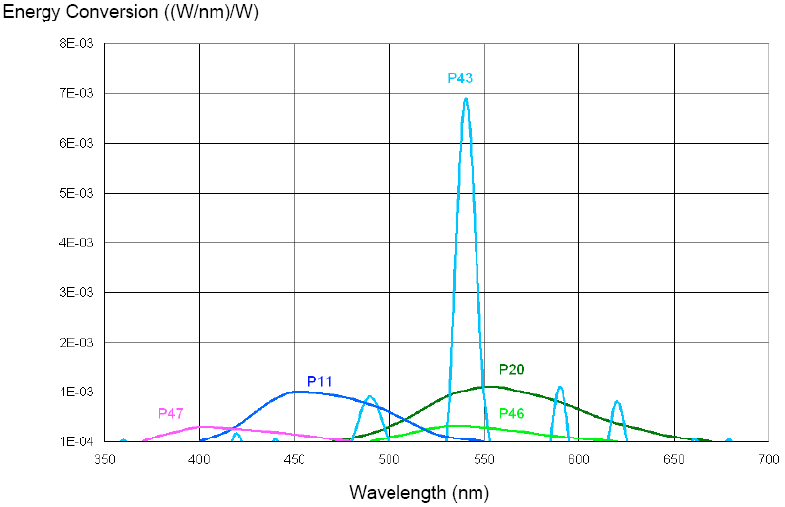}
\caption{Light emission spectra of typical phosphor materials}
\label{fig:phosphors}
\end{figure}

Special types of materials have been developed in order to obtain
large light yields, these are usually referred to as phosphors and were
primarily developed for the cathode ray tube industry. \Tref{tab:phosphors} gives
the compositions and the decay times for a few widely used phosphor
types while \Fref{fig:phosphors} shows the relative light emission
spectra and emission efficiencies.

\begin{table}
\caption{Composition and decay times of typical phosphor materials}
\label{tab:phosphors}
\centering
\begin{tabular}{|c|c|c|c|}
\hline
\textbf{Type} & \textbf{Composition} & \multicolumn{2}{c|}{\textbf{Decay time (of light intensity)}} \\\hline
~ & ~ & \textbf{From 90\% to 10\%} & \textbf{From 10\% to 1\%} \\\hline
\textbf{P 43} & $\rm Gd_2 O_2 S\colon Tb$ & 1\Ums & 1.6\Ums \\\hline
\textbf{P 46} & $\rm Y_3 Al_5 O_12\colon Ce$ & 300\Uns & 90\Uus \\ \hline
\textbf{P 47} & $\rm Y_2 Si O_5\colon Ce,Tb$ & 100\Uns & 2.9\Uus \\ \hline
\textbf{P 20} & $\rm (Zn, Cd)S\colon Ag$ & 4\Ums & 55\Ums \\\hline
\textbf{P 11} & $\rm ZnS\colon Ag$ & 3\Ums & 37\Ums \\
\hline
\end{tabular}
\end{table}

These phosphor materials have very large light yields, but can only be
used as thin coatings on rigid substrates and can be easily
damaged. Typical grain sizes for the coatings are of the order of the
micron, this represents the ultimate resolution limit for the imaging device. These materials can only be used for very low
intensity and low energy beams.

Ceramics materials, glasses, and crystals are much more frequently used in high--energy accelerators.
The light yields are usually lower than for the phosphors, but the radiation hardness and thermo-mechanical properties
are much better. A typical material widely used in accelerators is aluminium oxide, also known as \emph{alumina}, doped with chromium (\emph{Chromox} $\rm Al_2O_3\colon Cr$). This material is particularly robust and is well suited for the fabrication of beam observation screens.
Another widely used material that associates high light yield, high spatial resolution and good thermo-mechanical properties is YAG $Y_3 Al_5 O{12}$. In recent years a big effort has been made in order to improve the resolution of scintillating screens driven by the free--electron--laser light sources community. More details on scintillation and screens can be found in Ref. \cite{ariesapril2019}

\subsubsection{\v{C}erenkov radiation}
\v{C}erenkov radiation is a form of electromagnetic radiation emitted by charged particles while moving inside a material. The emission process is linked to the polarisation of the material and only takes place when the particles' velocity is larger than the phase velocity of the light in that material. The threshold velocity is given by the formula
\begin{equation}
\beta_t = \frac{1}{n(\omega)} ,
\label{eq:cerenkovthreshold}
\end{equation}
where $\beta_t=v_t/c$ is the threshold value of the relativistic speed factor and $n(\omega)$ is the wavelength--dependent refraction index of the material.

\begin{figure}[htb]
\centering\includegraphics[width=0.3\linewidth]{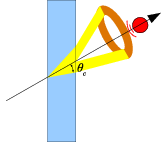}
\caption{\v{C}erenkov radiation is emitted by a particle travelling faster than the phase velocity of light in a material. The \v{C}erenkov photons have a well defined angular distribution.}
\label{fig:cerenkov}
\end{figure}

As the radiation is the result of the superposition of consecutive wave-fronts, the emission has a very particular angular distribution (see \Fref{fig:cerenkov}) with the photons emitted on the surface of a cone whose axis lies along the direction of motion of the particle and with aperture angle given by
\begin{equation}
\cos\theta_c=\frac{1}{n(\omega)\beta} .
\label{eq:cerencovangle}
\end{equation}

The emission spectrum of the radiation is given by
\begin{equation}
\frac{d^2N}{dx d\omega}= \frac{\alpha z^2}{c}\sin^2\theta_c(\omega)=\frac{\alpha z^2}{c} \left( 1- \frac{1}{\beta^2 n^2(\omega)} \right) .
\label{eq:cerencovspectra} 
\end{equation}

\begin{figure}[htb]
\centering\includegraphics[width=0.7\linewidth]{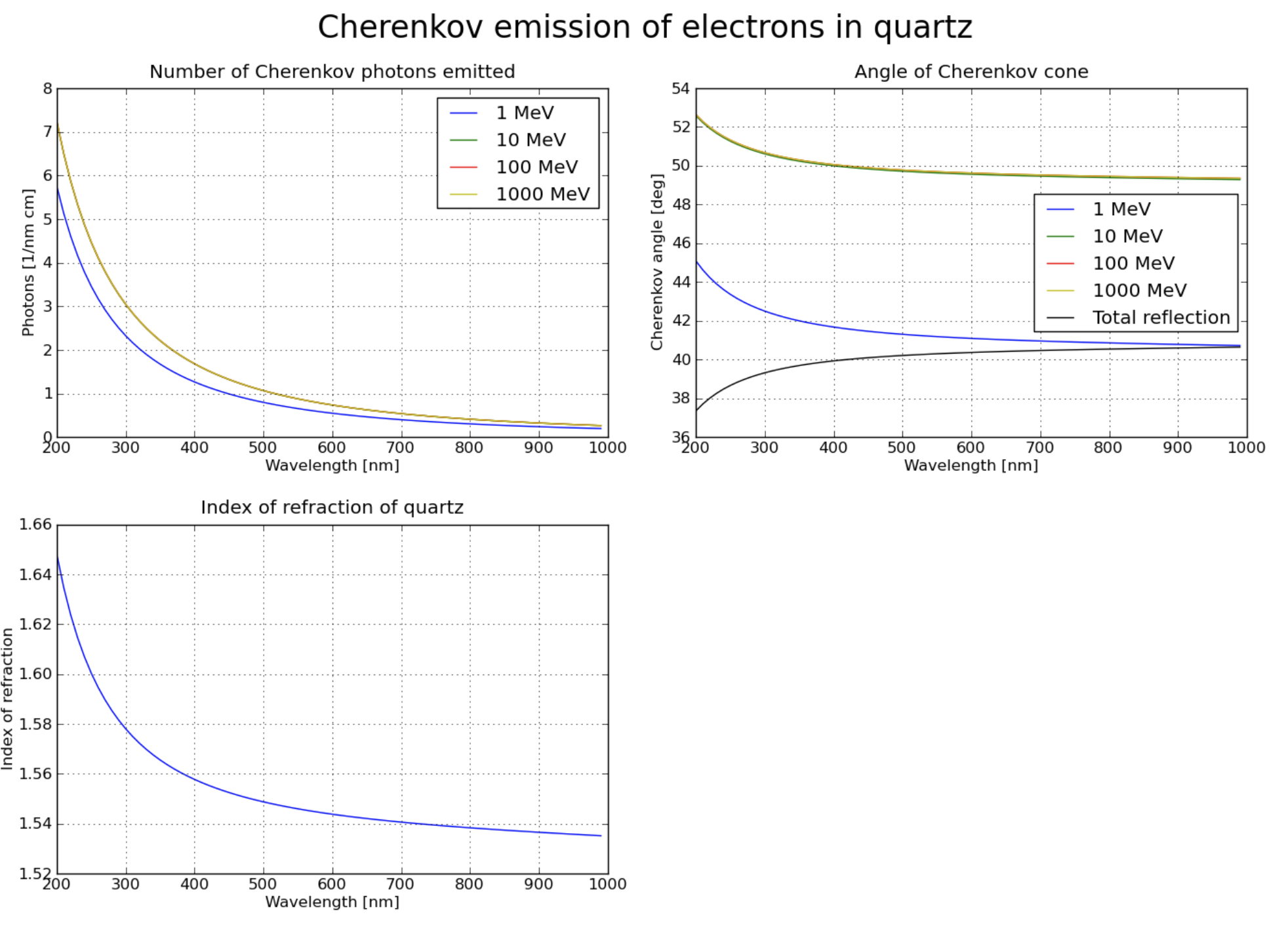}
\caption{\v{C}erenkov radiation emission by electrons of different energies in quartz. Emission intensity (top--left), emission cone angle (top--right) and index of refraction of quartz (bottom--left)}
\label{fig:cerenkovinquartz}
\end{figure}

As an example of the intensity of the \v{C}erenkov radiation, in the wavelength range $\lambda\in[400, 500]\Unm$, 10~photons are generated by a relativistic electron ($\beta\sim 1$) crossing a 1\Umm{} thick quartz plate ($n_{\rm quartz}= 1.46$, $\theta_c\simeq47^\circ$).
Because of the weak emission, relatively thick \v{C}erenkov radiators are required for beam imaging, this results in a limited spatial resolution. The light distribution at the exit surface consists of an ellipse obtained by the intersection of the \v{C}erenkov emission cone of the particle at the entrance in the material and the plane of the exit surface.
In diagnostic systems the angles between the radiator, the beam, and the detector need to be precisely aligned, for this reason it is important to take into consideration the refraction of the light at the exit surface.
\Figure[b]~\ref{fig:cerenkovinquartz} shows the property of the \v{C}erenkov radiation emitted by electrons of varying energy in quartz.

\subsubsection{Optical Transition Radiation (OTR)}
Optical Transition Radiation, also known as OTR, is a form of electromagnetic radiation emitted by a charged particle when crossing a boundary between materials with different dielectric properties. In most practical cases this boundary consists in the separation between the vacuum of the beam pipe and a metallic foil. OTR has many similarities with bremsstrahlung; the primary charged particle induces a mirror charge on the surface of the foil travelling in the opposite direction. When the primary charge enters the material these two charges neutralise. The effect is the same as if the mirror charge had come to a sudden stop. This radiation is called the \emph{backward emission} because it is emitted in the direction opposite to the direction of the particle; more precisely, the radiation is emitted along the direction of the specular reflection of the beam on the surface.

Similarly radiation is emitted when the primary particle exits the foil, in this case it is the sudden appearance of the real charge that emits the radiation called the \emph{forward emission}, always emitted in the direction of the particle independently of the angle between the particle and the surface.

OTR has a complicated angular distribution, a simplified description for $\gamma \gg 1$ of the case of the boundary between vacuum and a material with infinite $\varepsilon_r$ it is  given by the formula
\begin{equation}
\frac{d^2W}{d\Omega d\omega}\approx\frac{Nq^2}{\pi^2c}\left(\frac{\theta}{\gamma^{-2}+\theta^2}\right) .
\end{equation}

\begin{figure}[htb]
\centering
\begin{tabular}{c @{\hspace{1cm}} c}
\includegraphics[width=0.3\linewidth]{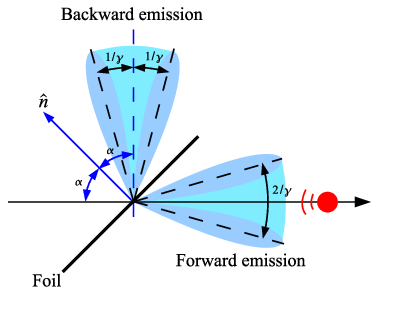} &
\includegraphics[width=0.3\linewidth]{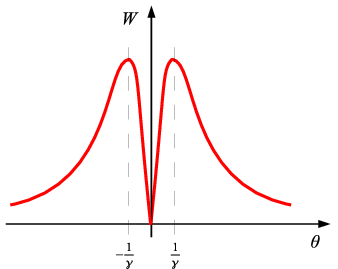} \\
a) & b) \\
\end{tabular}
\caption{OTR radiation is emitted by a particle crossing a boundary between materials with different dielectric properties (a). Angular distribution of the OTR photons (b).}
\label{fig:otr}
\end{figure}

The angular distribution corresponds to a thick walled cone of aperture angle $1/\gamma$ and the emission is characterised by a wide flat spectrum. For practical cases of metallic foils in vacuum (or air) one can still use this formula as the $\varepsilon_r$ of metals is very large for frequencies up to the plasma frequency of the material $\omega_p$ (which is linked to the density of electrons in the material).
\Figure[b]~\ref{fig:otr}(a) shows the geometry of the emission while \Fref{fig:otr}(b) shows the angular distribution, note the maximum at $1/\gamma$. 

In optical transition radiation the thickness of the radiator has no influence on the emitted radiation, it is thus possible to use very thin foils as radiators so that the perturbation to the beam is minimal. Similarly foils of materials with good thermo-mechanical properties can be used for the imaging of high density beams.

Another interesting characteristic of this type of radiation is that it is radially polarised, \ie the polarisation vector of the photons lies in the plane defined by the beam path (or its specular reflection for the backward emission) and the direction of the photons.
The angular distribution of OTR has a strong dependency on the energy of the particle and more precisely on the relativistic factor $\gamma$. The intensity of the radiation is also influenced by the energy in a complicated way. The complete model describing this type of radiation can be found in Refs. \cite{Ginzburg:306834} and \cite{Wartski:889093}.
A simple description of the dependency of the total OTR power as a function of the particle energy is given by
\begin{equation}
W\propto \left\{ 
\label{eq:otrpower}
\begin{array}{c c}
\beta^2 & \beta \ll 1\\
\ln{2\gamma} & \gamma \gg 1
\end{array}
\right. .
\end{equation}

For the wavelength range $\lambda \in [400, 600]\Unm$, an accurate calculation using the complete OTR model predicts $\sim$0.3~OTR photons per electron for 50\UMeV{} electrons ($\beta=1$, $\gamma=98.8$) and $\sim$0.001~photons per electron for 100\UkeV{} electrons ($\beta=0.55$, $\gamma=1.2$).

For the \emph{backward} emission there is one more important issue. The emission process can be seen as if the photons are created in the direction of the primary particle just before the particle crosses the surface of the foil and are then immediately reflected by the surface of the foil. In the reflection mechanism the properties of the surface are essential in defining the intensity, the spectrum, and the direction of the out coming photons. If the surface is not perfectly flat this will distort the particular angular distribution, in the extreme case of a very rough surface the backward emission can become isotropic. The radiation spectrum is also affected as there is a cut-off frequency at which the surface does not reflect any more, generally the resulting spectrum will be given by the convolution of the \emph{primary} radiation and the wavelength-dependent reflection coefficient.

Another important aspect of OTR is the coherent emission. When the beam is bunched in very short packets like in free electron laser LINACs, the particles emit with the same phase and the intensity of the emission does no longer scale linearly with the beam intensity. In the extreme case of a very short bunch the total radiation intensity would be proportional to the square of the number of particles $N^2$ in the bunch instead of $N$ see Ref. \cite{Loosy:2008zz}. For this reason OTR can not be used in certain cases. The free electron laser community had to revert to using scintillating screens in order to overcome this issue.

\subsubsection{Synchrotron radiation}
An effect that is not properly an interaction of charged particles with matter, but is very important in transverse profile measurements, is synchrotron radiation. 
From classical electrodynamic theory we know that when a charged particle is accelerated it emits electromagnetic radiation, \emph{bremsstrahlung} is probably the most widely known effect of this type. In the \emph{bremsstrahlung} case the radiation is emitted as a consequence of the charge being decelerated (\ie reduction of energy), there is, however, another type of acceleration that does not imply a change in the energy of the particle, but a change of direction.

The radiation emitted by a charged particle when its trajectory is deviated without changing its energy is known as \emph{synchrotron radiation} or \emph{synchrotron light}. As the name suggests this effect was initially observed in the bending magnets of relativistic circular machines, namely electron synchrotrons, but is in fact emitted any time the trajectory of a charged particle is modified. In accelerators the magnetic field of the bending magnets is used to force the trajectory of the particles into a circular orbit, as a by-product synchrotron radiation is emitted. This radiation can be exploited for scientific purposes like in synchrotron light sources, or can be a nuisance since the energy carried by this radiation is subtracted from the energy of the particles.

\begin{figure}[htb]
\centering
\begin{tabular}{m{0.5\linewidth} m{0.3\linewidth}}
\includegraphics[width=\linewidth]{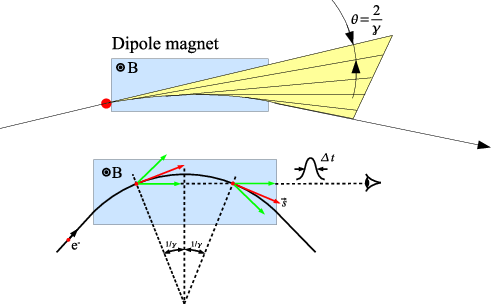} &
\includegraphics[width=\linewidth]{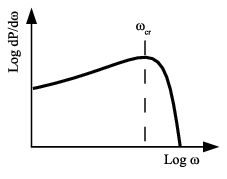} \\
\begin{center} a) \end{center} & \begin{center} b) \end{center} \\
\end{tabular}
\caption{A charged particle emits synchrotron light while travelling inside a bending magnet (a). Spectrum of the radiation (b).}
\label{fig:synchdipole}
\end{figure}

The fraction of energy lost by the beam via synchrotron radiation can be very important and constitutes a huge limitation on the maximum energy of electron synchrotrons. The Large Electron--Positron collider (LEP) at CERN, for example, accelerated counter rotating electrons and positrons up to 100\UGeV{} ($\gamma \approx 2\,10^5$) and the RF system had to restore at each turn about 3\% of the energy of the beam that was lost by synchrotron radiation. The total irradiated power was of the order of 10\UW[M]{} \cite{Jowett:359896}. In contrast the Large Hadron Collider (LHC) with the same circumference. a proton beam energy of $7\UTeV$ ($\gamma \approx 7\,10^3$) and a beam current about 100 times that of LEP only emits about $2\UW[k]{}$.

\begin{figure}[htb]
\centering\includegraphics[width=0.5\linewidth]{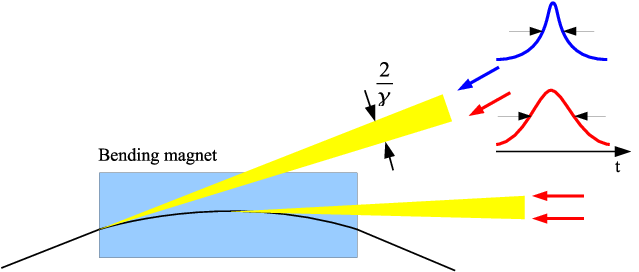}
\caption{The pulse length of the synchrotron light seen by an observer looking at the entrance/exit edge of the magnet is shorter than the pulse seen by an observer looking inside the magnet}
\label{fig:synchpulse}
\end{figure}

The power emitted by a charged particle in the form of synchrotron radiation inside a bending magnet is given by \cite{Hofmann:133993}
\begin{equation}
P=\frac{1}{4\pi\varepsilon_0}\frac{2}{3}\frac{c e^2\gamma^4}{\rho^2} .
\label{eq:synchrotronpower}
\end{equation}
The photons are emitted with a strongly forward peaked distribution (arising from the relativistic transformation from the particle reference frame to the laboratory reference frame), the aperture of this distribution is approximately given by $2/\gamma$. As a particle travels around an accelerator it will continuously enter and exit bending magnets. The radiation emission will therefore start at the entrance of the magnet, remain constant while inside the magnet and finally stop at the exit of the magnet. An observer looking from above would see the light continuously emitted as a fan (like a car driving round a bend at night) see \Fref{fig:synchdipole}. On the other hand, an observer looking into the beam pipe would only see a flash whose duration corresponds to the difference in travel time between the photons and the particle during the time it takes the particle to be deviated by the angle $2/\gamma$, see \Fref{fig:synchdipole}. If the same observer would look at the entrance/exit edge of the magnet (blue arrows on \Fref{fig:synchpulse}) he would see an even shorter pulse.

The spectrum of the synchrotron radiation can be obtained by Fourier-transforming the pulse shape. The pulse length inside the magnet is given by
\begin{equation}
\tau=\frac{2\rho}{\gamma\beta c}-2\sin\left(\frac{1}{\gamma}\right)\frac{\rho}{c}\approx \frac{4}{3}\frac{\rho}{c\gamma^3} ,
\label{eq:synchpulse}
\end{equation}
with the typical frequency around
\begin{equation}
f_{\text{typ}}\sim \frac{1}{\tau} \approx \frac{c\gamma^3}{\rho} .
\label{eq:synchfreq}
\end{equation}

An important parameter of synchrotron radiation is the critical frequency $\omega_{cr}$ and it is defined so that one half of the total radiation power is emitted at frequencies below $\omega_{cr}$ and the other half above
\begin{equation}
\omega_{cr}= \frac{3}{2} \frac{c\gamma^3}{\rho} .
\label{eq:synchcritfreq}
\end{equation}

In beam diagnostics it is often preferred to build devices operating in the visible range for simplicity, in some conditions, however, the emission spectrum of synchrotron radiation from a bending magnet does not cover this range, in these cases the so--called \emph{edge radiation} can be useful. The concept is that, from fundamental Fourier transform properties, shorter pulses correspond to wider spectra and thus edge radiation extends to lower (and higher) wavelengths compared to simple \emph{dipole radiation}; this trick is often used in high--energy proton machines where the synchrotron radiation just starts to be exploitable for diagnostic purposes.

\begin{figure}[htb]
\centering\includegraphics[width=0.5\linewidth]{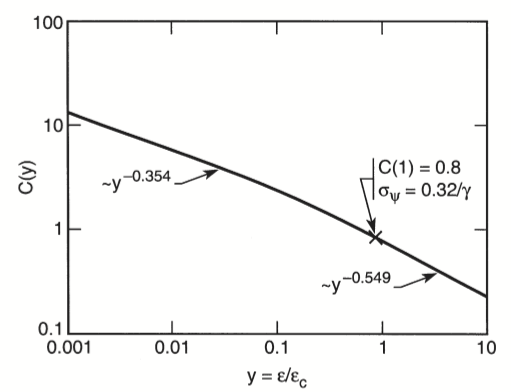}
\caption{Synchrotron radiation opening angle variations as function of $\omega/\omega_{cr}$}
\label{fig:synchlightopeningangle}
\end{figure}

In case we want to observe the synchrotron radiation at wavelengths far from the critical frequency $\omega_{cr}$ we have to take into account that the emission cone will have an aperture quite different from the $2/\gamma$ mentioned earlier
\begin{equation}
\sigma_\theta = \frac{2}{\gamma \sqrt{2 \pi}} C(y) ,
\label{eq:synchopeningangle}
\end{equation}
where $C(y)$ is a tabulated function of $y=\omega/\omega_{cr}$ that can be seen in Fig.~\ref{fig:synchlightopeningangle}.

If the bending magnet is replaced by a sequence of dipole magnets with alternating polarities, it is possible to produce synchrotron radiation without significantly modifying the trajectory of the particle, typically a device with three poles is used and takes the name of \emph{wiggler} magnet. This device is usually used where the important aspect is to force the beam to produce synchrotron light with no particular interest in the properties of the radiation. If the device consists of a large number of regular periods it takes the name of \emph{undulator}; the particles instead of being just deviated will oscillate around the central trajectory. The resulting radiation will have a much narrower spectrum centred on a well defined wavelength defined by the undulator period length
\begin{equation}
\lambda = \frac{\lambda_u}{2\gamma^2} ,
\label{eq:synchunulatorlambda}
\end{equation}
and the emitted power is
\begin{equation}
W\propto B_0^2\gamma^2 ,
\label{eq:synchunulatorpower}
\end{equation}
where $B_0$ is the absolute magnetic field in the centre of the pole.

\subsubsection{Inverse Compton scattering}
\begin{figure}[htb]
\centering\includegraphics[width=0.3\linewidth]{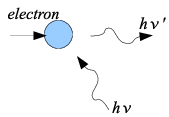}
\caption{A low--energy photon interacts with a high--energy electron. A fraction of the energy of the electron is transferred to the photon in the process.}
\label{fig:compton}
\end{figure}

In the theory developed by Compton \cite{PhysRev.22.409}\cite{PhysRev.24.168}, the scattering of a high energy photon on an electron at rest is described. The case where the energy of the electron is much larger than the energy of  the photon is usually referred to as \emph{inverse Compton scattering} (\Fref{fig:compton}).

In order to understand inverse Compton scattering, it is sufficient to move to the reference frame of the high energy electron; in this reference frame what is observed is just the well known Compton scattering with the electron at rest and the photon having a much higher energy than in the laboratory frame. After this change of reference the interaction can be calculated with the existing detailed mathematical model of Compton scattering. After this it is sufficient to re-transform the resulting particles back to the laboratory reference frame, in this process the photon will increase again its energy (gamma boost).

In inverse Compton scattering the energy of the photon gets a boost of the order of $\gamma^2$ at the expense of the energy of the electron, and the direction of the emerging photon is peaked around the direction of the electron. In Compton scattering there is a strong correlation between the energy lost by the photon and the scattering angle. The same happens in inverse Compton scattering with the difference that this time the photon gains energy and the relation between angle and energy is complicated by the two reference frame transformations.

In beam diagnostics the electron usually has high energy (typically in the \UGeV{} range) and the photon is in the range from infra-red (IR) to ultra-violet (UV) ($\sim$1 to 4\UeV{}) where powerful lasers are available.

The total cross section for Compton scattering, described by the Klein--Nishina formula, is rather small, of the order of $10^{-25}\Ucm^2$ for electrons \cite{Fargion:305581}, but with modern lasers it possible to obtain measurable signals for electron/photon (or positron/photon) collisions. Although this effect exists for proton beams as well (usually called Compton scattering on protons), the cross section is much smaller as it scales with the square of the rest mass and it is thus not possible to obtain sufficient signals.

\subsubsection{Photo dissociation}
Another interesting phenomenon concerning the use of lasers in beam diagnostics is photo--dissociation or photo--neutralization. In this case photons are used to detach the extra electron from negatively charged hydrogen ions (H${}^-$). This process can be facilitated by external electric or magnetic fields that if opportunely applied reduce further the already small ionization potential.

\begin{figure}[htb]
\centering\includegraphics[width=0.7\linewidth]{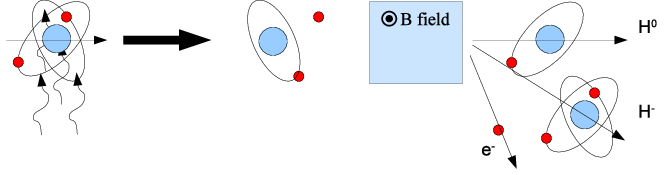}
\caption{A laser beam is used to free the extra electron on an H${}^-$ ion resulting in a neutral hydrogen atom and a free electron travelling at the same speed. A dipole magnet is used to separate the three species emerging from the interaction.}
\label{fig:photodissociation}
\end{figure}

This process has attracted a lot of interest recently as most new developments of high current proton linear accelerators favour the use of H${}^-$ beams. The reason for this is that the injection inside a downstream circular machine can be done by stripping the two electrons instead of using septa and kickers allowing what is called \emph{phase--space painting} resulting in much more brilliant beams. \Figure[b]~\ref{fig:photodissociation} shows a schematic of photo-neutralization and how the different species can be separated for detection.

\section{Sampling techniques}
We have just seen the different processes that can be used for the detection of particles and the sampling of distributions. Apart from intercepting vs. non--intercepting, there is one additional distinction that can be made on sampling techniques, and precisely between one--dimensional sampling and two--dimensional sampling.

While two--dimensional sampling allows the direct sampling of the transverse space, the 1D sampling only allows the acquisition of projections. To express this difference more clearly, think of a beam where coupling deforms the distribution of particles in the transverse geometrical space into a tilted ellipse (i.e., an ellipse whose major and minor axes are not aligned with the $x$ and $y$ axes). If we use a 2D sampling device, \ie we take a picture of the beam, we immediately observe it. On the contrary, if we use only 1D sampling devices, \ie we acquire projections, we will not notice that the ellipse is tilted, and this even acquiring $x$ and $y$ projections simultaneously.

In the particular case presented above, the problem could be solved by acquiring three one--dimension profiles, one along $x$, one along $y$, and the third at $45^{\circ}$. In case of an arbitrary transverse distribution, a complex system with a large number of 1D projections can eventually disclose the full distribution (tomography). In many machines, however, especially high--energy circular accelerators, the shape of the distribution is known and the measurement of the $x$ and $y$ profiles is sufficient to define the scaling parameters (usually with a fit of the measured data to a model).  

In the one--dimensional sampling we have
\begin{itemize}
\item Wire scanners
\item Wire grids
\item Rest gas ionization monitors
\item Laser wire scanners
\end{itemize}

While in the two dimensional sampling we have
\begin{itemize}
\item Screens and radiators
\item Synchrotron radiation
\end{itemize}

\subsection{Sampling projections}
\subsubsection{Wire scanners}
\begin{figure}[htb]
\centering\includegraphics[width=0.7\linewidth]{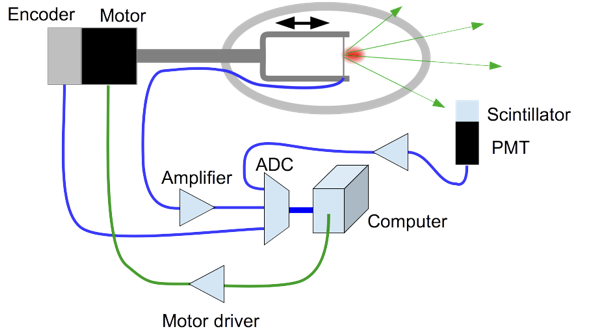}
\caption{A thin wire is scanned through the particle beam while the secondary emission current, the signal from a calorimeter downstream, and the signal of the motor encoder are acquired simultaneously. Plotting either of the SEM or PMT signals against the encoder gives the beam profile. }
\label{fig:wirescanner}
\end{figure}

\begin{figure}[htb]
\centering\includegraphics[width=0.7\linewidth]{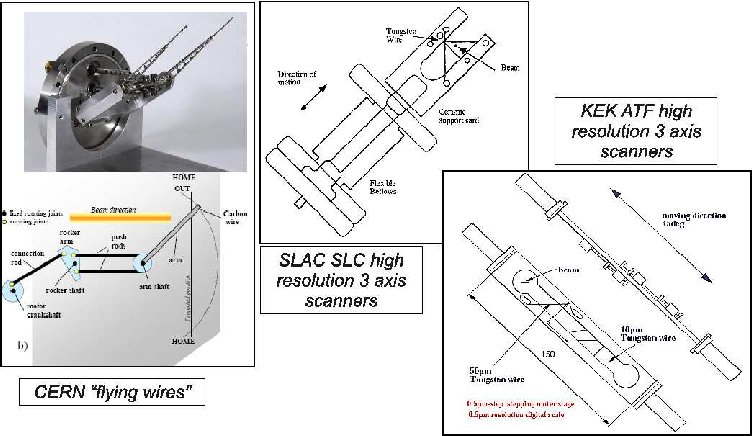}
\caption{Different types of rotative and linear wire scanners}
\label{fig:wirescanner2}
\end{figure}

The principle of the wire scanner is rather simple and consists of moving a wire across the beam while monitoring a signal proportional to the number of particles interacting with the beam, see \Fref{fig:wirescanner} and \Fref{fig:wirescanner2}. Certainly the wire has to be placed inside the vacuum chamber increasing the complexity of the system. The signal observed is usually either the secondary emission current (SEM) from the wire, or the flux of high energy secondary particles downstream of the wire. The high energy secondaries are often detected with scintillators coupled to photomultiplier tubes (PMT). Neutral optical filters could be interposed between the scintillators and the PMT in order to avoid saturating the PMT and increase the dynamic range.

The precision of the wire scanner is dominated by two aspects: the precision with which the wire is positioned and the precision of the acquired signal.
The major problem in building a wire scanner is to design a solid and stable mechanical support for the wire and an accurate mechanism for the movement. Frequently some sort of encoder or resolver is used to read the position of the wire support (fork).

The speed of the movement mechanism varies over a very large range depending on the application. On linacs or transfer lines where the beam consists of short pulses at low repetition rate, a high speed device is not required as the measurement is performed by acquiring one sample on each beam pulse (usually < 100\UHz{}) leaving plenty of time to move the wire from one pulse to the next, for example, 500\Uum{} steps at 100\UHz{} correspond to a speed of $5\Ucm/\UsZ$. On the other hand, in circular accelerators where either the beam intensity is elevated or the energy of the machine is rapidly increasing a fast scan is essential (to reduce wire heating in the first, and to sample the whole profile at the same energy in the second). In these cases wire speeds of $\sim 20\Um/\UsZ$ have been achieved (typical revolution frequencies in circular accelerators exceed 1\UMHz{}).

Over heating of the wire and consequent wire damage is the most common problem with wire scanners, see, for example, Ref. \cite{Sapinski:1123363} for a detailed description.

Although there are deconvolution techniques that can help reduce the error introduced by the finite size of the wire, it is better to keep the wire diameter much smaller than the size of the beam to be measured. For this reason very thin wires are used, in the range between $\sim 5\Uum$ to $\sim 50\Uum$ (of course thicker wires are possible for large beams). The mechanical constraints imposed by the deformation or damage of the wires is the main limitation for the speed. The deformation of the mechanics and of the wire are also the main source of error in fast wire scanners, for this reason all high resolution scanners have slow movements \cite{hayano-2000-000821}\cite{Arduini:793452}. 

The SEM signal is typically used with low energy beams as in this case no energetic secondary particles are generated; this signal tends to be quite small and requires care in the acquisition. A serious problem with the detection of the secondary emission is the fact that when the wire is heated above 1000\UDC{} by the beam it starts emitting electrons by thermionic emission perturbing the measurement of the SEM current.

The signal from high energy secondaries is typically large due to the high gain of the scintillator/photo tube detector. On the other hand, beam losses can pollute the signal and, more importantly, due to the geometry of the detector and of the beam line, the signal induced in the detector may depend on the position of the wire and direction of the particles, introducing distortions and aberrations in the profiles.

The detection of the high energy secondaries is often accomplished by mean of photomultipliers. It worth mentioning that the gain of a photomultiplier depends on the voltage potential applied to the various dynodes. The different bias voltages are usually obtained by resistive dividers inside the base and the total current draw at the anode affect this partition. It is very important to ensure that the output level of the photomultiplier is well inside the linear region where the anode current does not perturb the gain.

The dimension of the wire has a direct impact on the signal strengths for both techniques; in the case of SEM the signal is proportional to the radius of the wire (SEM is a surface effect only) while for the high energy secondaries the signal is proportional to the square of the radius as it is a volume effect.

\begin{figure}[htb]
\centering\includegraphics[width=0.5\linewidth]{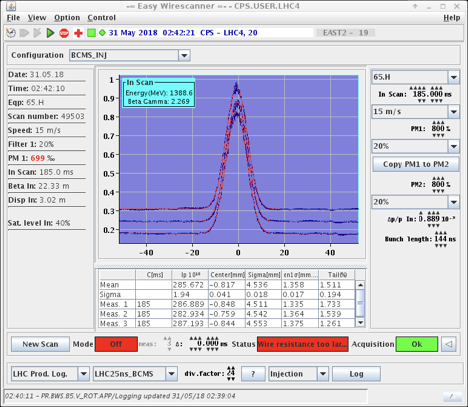}
\caption{Example of beam profiles acquired with a wire scanner}
\label{fig:wirescannergui}
\end{figure}

Figure~\ref{fig:wirescannergui} shows an example of beam profiles acquired with a wire scanner in the CERN PS accelerator.

\begin{figure}[htb]
\centering\includegraphics[width=0.5\linewidth]{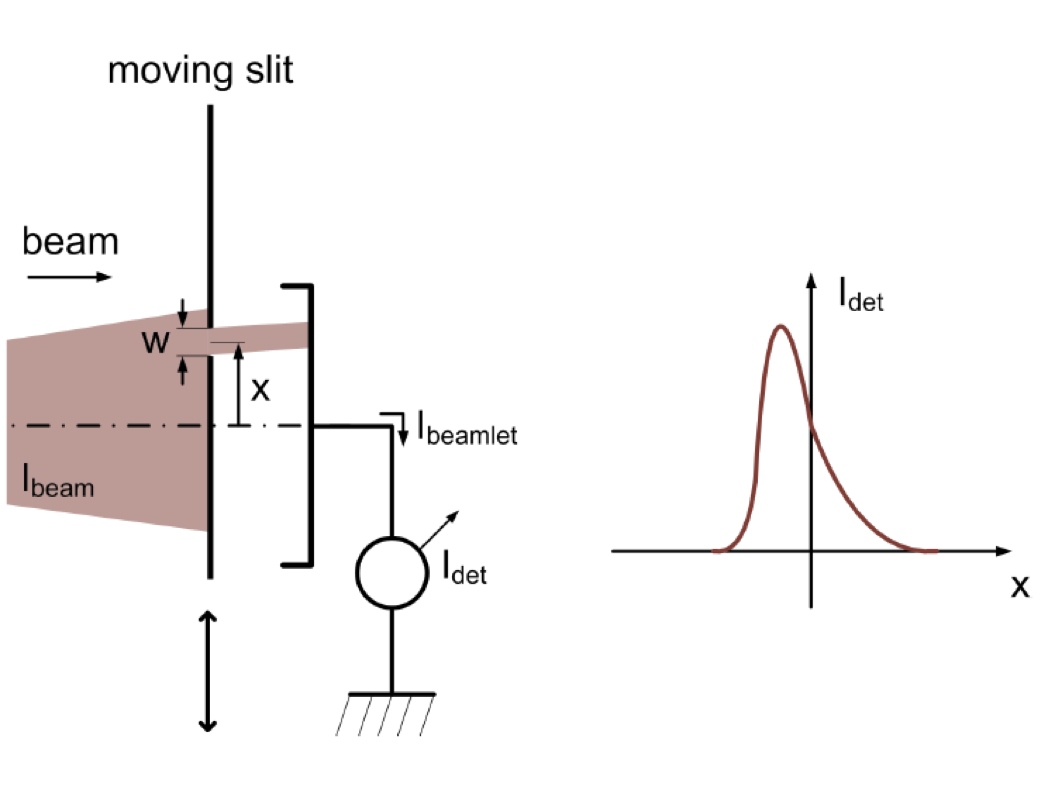}
\caption{Principle of the slit scanner. A thin slit cut into a solid blade is scanned across the beam. At the same time the beam current downstream of the blade is recorded and plotted against the slit position.}
\label{fig:slitscanner}
\end{figure}

The slit scanner can be consider as a special case of the wire scanner. It is composed of a solid blade in which a thin slit is cut. In this device, instead of exploting the interaction between the this wire and the beam, the blade is used to stop the beam and only let through particles inside the slit. The beam profile is obtained by acquiring the \emph{beamlet} current downstream of the blade, using a Faraday cup for example, as function of the slit position as can be seen in Fig.~\ref{fig:slitscanner}. Slit scanners are usually employed for low energy and low intensity beams and use sensitive current sensor. 

\subsubsection{SEM grids (wire harps)}
\begin{figure}[htb]
\centering
\begin{tabular}{m{0.5\linewidth} m{0.5\linewidth}}
\includegraphics[width=\linewidth]{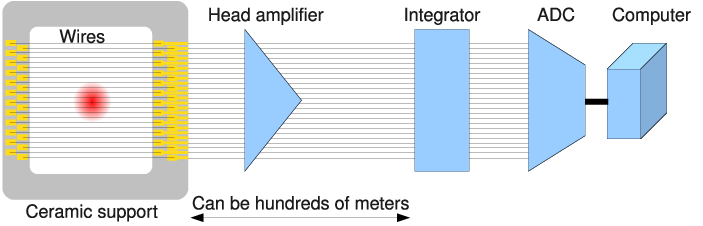} &
\includegraphics[width=\linewidth]{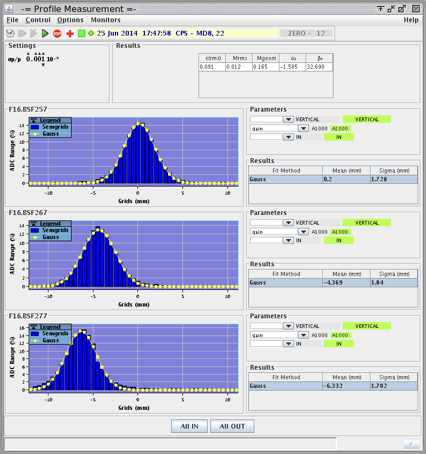}\\
\end{tabular}
\caption{Schematics of a SEM grid system (left) and an example of profiles acquired simultaneously at three different locations along an emittance measurement line (right)}
\label{fig:semgrid}
\end{figure}

Wire harps are devices where a large number of, parallel, fixed wires or strips are placed on the beam path, \Fref{fig:semgrid}. In this case the high energy secondaries generated in the interactions between the beam and the wires can not be used as it is impossible to distinguish from which wire the particles have been generated. The acquisition of the secondary emission current from each individual wire is then the only possibility.

The advantage of a wire harp over the wire scanner is that it allows single--shot acquisitions and placing several harps one after the other allows single--shot measurement of the different planes ($x$ and $y$) and/or the acquisition of the same plane at different locations (a technique used to measure the beam emittance). Another advantage over the wire scanner is the absence of moving parts, apart from a slow actuator used to insert and retract the device if needed.
The drawback is the need for a complicated acquisition system with many channels and small signals and a limit on the spatial resolution as the wire spacing can hardly be reduced to less than a few hundred micrometres.

Typically the acquisition chain is composed of a head amplifier that sits as near as possible to the grid, often an area with high radiation, followed by an integrator or signal conditioning circuit that sits far away in a safe counting room, and finally a computer controlled ADC. This scheme requires one signal cable per wire over the distance from the device to the counting room and the cabling can easily become the most expensive part of the system. In order to reduce the costs, techniques have been developed for sending the signals over twisted--pair cables. Although this solution can cut the cabling costs by a large factor it also introduces limitations on bandwidth and cross--talk.

The number of wires in segmented detectors like the SEM grids varies from system to system. As a rule of thumb for Gaussian beams one can consider the maximum wire spacing to be half the beam $\sigma$ and the detector size at least four $\sigma$'s which means a minimum of eight wires. This rule is valid for a perfect Gaussian beam  well centred. In reality a denser mesh and a larger detector is better. Typical real systems range from 16 to 48 wires.

\begin{figure}[htb]
\centering\includegraphics[width=0.7\linewidth]{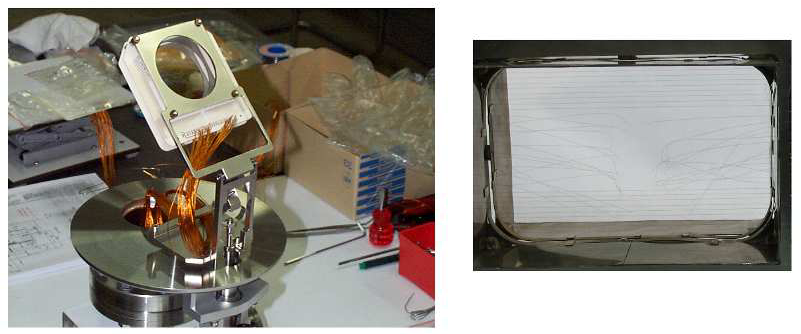}
\caption{Pictures of a SEM grid assembly (left) and of damaged 50\Uum{} tungsten wires (right)}
\label{fig:semgrid2}
\end{figure}

If the signal--to--noise allows, it is possible to acquire the signals from the wires at high speed and observe the evolution of the profiles inside a beam pulse. Acquisition frequencies up to 100\UMHz{} have been achieved.
As with wire scanners the damaging of the wires is the most important failure and having many wires permanently in the beam increases the risk, \Fref{fig:semgrid2}. For the wire grids even a single broken wire can prevent the whole system from functioning correctly as the damaged wire can short--circuit several neighbouring intact wires rendering them unusable.

One important detail for wire grids consists in avoiding that secondary electrons generated on one wire get reabsorbed by another wire as this would generate cross--talk between channels. To avoid this problem, clearing electrodes are installed around the device to generate an electric field whose lines  pull the secondary electrons away from the wires.

Thanks to the huge progress of microelectronics in recent years, it is now possible to develop radiation tolerant Application Specific Integrated Circuits (ASIC) dedicated to the acquisition of the SEM grid signals. Similar circuits are common in large physics experiments (like ALICE and ATLAS at the LHC), see Ref.~\cite{Anghinolfi:2004dw} for example, but the technology was too complicated and expensive for small projects like a few SEM grids. Today the development and prototyping is much more affordable.

A dedicated ASIC embedded in the harp can increase the sensitivity, decrease the noise, improve the time resolution and reduce the cabling costs, in particular if the chip converts the signals to a digital stream of data. 

\subsubsection{Ionisation profile monitor}
Owing to the problem of wire damage, both the wire scanners and the wire grids can not be used in high intensity beams or in continuous monitoring in circular machines. In these cases non--intercepting devices are needed and the ionisation profile monitor has been developed precisely for these needs.

The ionisation profile monitor (IPM) is based on the interaction between the beam and the rest gas present in the vacuum chamber, even in the best vacuum there are still $\sim 10^{13}$~ions/cm$^3$. If the rest gas density is not sufficient it can be artificially increased locally with a small gas injection.

When the particles of the beam pass through the rest gas these ionise the atoms leaving behind a column of ions and electrons. The spatial distribution of these secondaries is the same as that of the distribution of the primary particles. By detecting the distribution of the secondaries, it is thus possible to reconstruct the distribution of the particles of the beam.

The detection of the secondaries is achieved by applying an electric field perpendicular to the beam trajectory and to the plane to be observed, this field will accelerate the ions and the electrons in opposite directions. At a certain point, well clear of the beam path, a detector is installed that intercepts the drifting particles and records their arrival position.

In fact there are two options for the IPM, detecting either the electrons or the ions; in principle by switching the electric field polarity one could switch from one particle to the other, in reality the systems are optimized for one case or the other. In most cases the electrons are observed and in this case a magnetic field parallel to the electric field is required. The role of this magnetic field is to reduce the transverse drift of the electrons due to their random initial velocity and the acceleration due to the space charge of the beam (radial electric field). The electrons will spiral around the magnetic field lines while being pushed towards the detector by the electric field.

The detector can be a simple grid of collecting electrodes or an optical system based on a multichannel plate and a phosphor screen. Recently silicon detectors, similar to those used in the tracking systems of the particle physics experiments, have bee used in IPMs. A good example of this new variant is the IPM developed for the CERN PS synchrotron~\cite{Storey:2018vuk}.

The first type of detector is similar to a SEM grid, but instead of the SEM current it is the current from the collection of the charges that is measured, like the anode current in a vacuum tube. A positive bias potential on the wires or strips is required in order to capture and retain the incoming low energy electrons and avoid SEM emission. In case of ions a negative voltage is applied and the signal is the superposition of charge collection and SEM.

\begin{figure}[htb]
\centering
\begin{tabular}{m{0.5\linewidth} m{0.3\linewidth}}
\includegraphics[width=\linewidth]{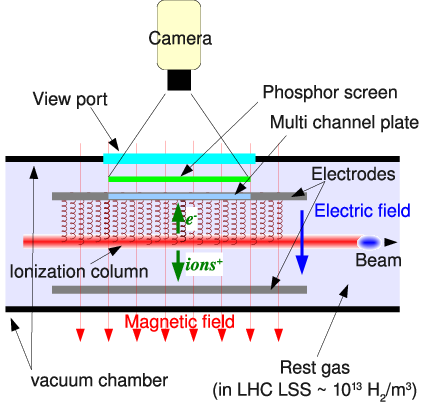}  &
\includegraphics[width=\linewidth]{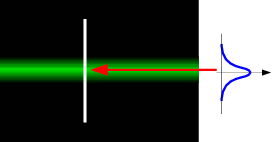} \\
\end{tabular}
\caption{Schematic of an IPM monitor (left) and the relative image observed on the electron detector (right)}
\label{fig:ipm}
\end{figure}

\begin{figure}[htb]
\centering\includegraphics[width=0.8\linewidth]{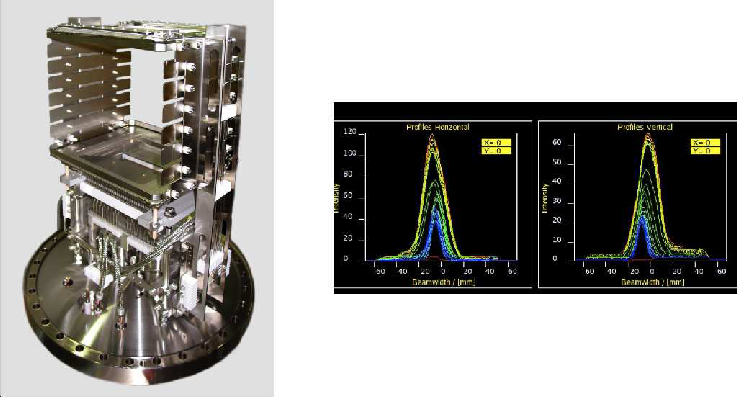}
\caption{Picture of an IPM monitor at GSI (left) and example of measurements (right)}
\label{fig:ipm3}
\end{figure}

The optical detector is based on the multichannel plate, an electron multiplier array that conserves the spatial information. The amplified electrons are then accelerated toward a phosphor screen by means of an electric field of a few kilovolts. Finally a video camera records the image formed on the screen  (a stripe). \Figure[b]~\ref{fig:ipm} shows a sketch of an IPM system and the image acquired by the video camera while \Fref{fig:ipm3} shows the picture of a real device and related measurements.

Ionization profile monitors often suffer from artefacts in the measurement, the most important being the tails arising from the transverse drift of the electrons or ions during their travel towards the detector not being completely prevented by the magnetic field. More information on IPMs can be found in Refs. \cite{Satou:1078651} and \cite{Forck:924577}.

\subsubsection{Laser wire scanner}
Another type of non--intercepting 1D profile monitor is the laser wire scanner. This device is based on the inverse Compton scattering (ICS) described before and is thus only available for electron and positron beams.

\begin{figure}[htb]
\centering
\begin{tabular}{m{0.59\linewidth} m{0.36\linewidth}}
\includegraphics[width=\linewidth]{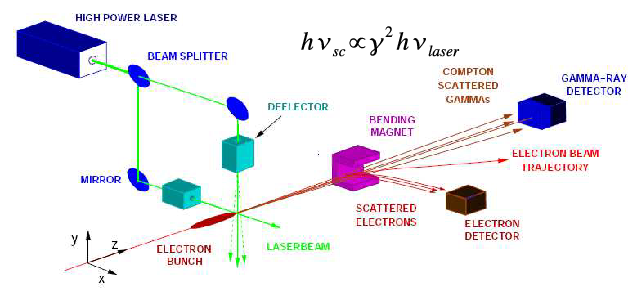} &
\includegraphics[width=\linewidth]{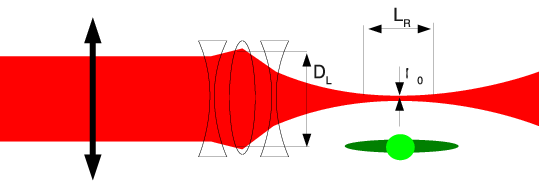} \\
\end{tabular}
\caption{Schematic of a laser wire scanner system (left) and detail of the laser focusing system (right)}
\label{fig:lws}
\end{figure}

The basic concept is quite simple and is depicted in \Fref{fig:lws}, a powerful, well focused laser (referred to as the laser wire) is scanned across the beam to be measured, as is done in a traditional wire scanner. The photons of the laser interact with the high energy electrons and create high energy X-rays or $\gamma$-rays, with an energy boost of the order of $\gamma^2$. A detector downstream detects the flux of those particles. By plotting the number of gammas against the position of the laser the profile of the beam is obtained.

In order to detect the generated gammas it is necessary to separate them from the primary electron beam, this is usually done by means of a dipole magnet. In circular machines this can be a bending magnet of the lattice, in linear machines or transfer lines a magnetic chicane may have to be added deliberately. When the energy of the particles is very high, it may be difficult to deflect them just for the purpose of detecting the gammas. In this case either the LWS device has to be installed in a location where the particles are already deviated from their linear trajectory for other reasons, or a different detection technique has to be used.

In the ICS process we have seen that a fraction of the energy of the particles is transferred to the photons. This fraction can be very high for ultrarelativistic beams. As a consequence the electrons that have interacted with the photons will have an energy much lower than the average energy of the beam, and are for this reason called degraded particles (degraded electrons). As the optics of the beam line is designed for the average beam energy, the degraded particles can not be transported very far and can be detected as beam losses, perhaps by sophisticated ad hoc calorimeters placed in optimized locations. In order to facilitate the collection of the degraded particles, additional magnetic systems may be introduced.

When a laser beam is focused on a small spot the divergence of the photons becomes important and the laser beam remains focused only over a short distance, limiting in practice the maximum usable length of our \emph{wire}, this represents a major difference between the laser wire and a real wire. This limitation depends on the size of the waist of the laser beam, the smaller the waist the shorter the focused length defined by the Rayleigh length $L_\text{R}$.
On the other hand, lasers can be focused to very small spots,  about one order of magnitude smaller than the thinnest wire and thus allow the measurement of very small beams.
 
The size of the waist of the laser is given by
\begin{equation}
\sigma_0=\frac{\lambda f}{D_\text{L}}= \lambda f/\# ,
\label{eq:lwswaist}
\end{equation}
where $\lambda$ is the wavelength of the laser, $f$ is the focal length of the lens (or lens system), and $D_\text{L}$ is the clear aperture of the lens system. The ratio $f/D_\text{L}$ is often referred as the \emph{stop number} in optics and denoted by $f/\#$.

The Rayleigh length represents the distance over which the laser spot size remains within $\sqrt{2}$ of $\sigma_0$ and is given by
\begin{equation}
L_\text{R}=\frac{2\pi\sigma_0^2}{\lambda} .
\label{eq:lwsraleigh}
\end{equation}

Laser wire systems usually require very powerful lasers as the cross section for the inverse Compton process is quite small. Typical lasers used for these applications deliver several millijoules of laser power in pulses of few picoseconds. The quality of the laser beam is also very important as it affects the smallest waist size that can be attained and is usually expressed in terms of the $M^2$ factor, $M^2=1$  denotes a pure monomode Gaussian beam.

\begin{figure}[htb]
\centering
\begin{tabular}{m{0.39\linewidth} m{0.59\linewidth}}
\includegraphics[width=\linewidth]{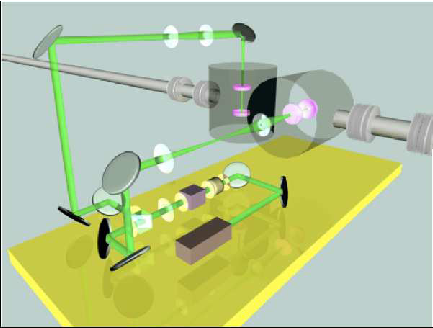} &
\includegraphics[width=\linewidth]{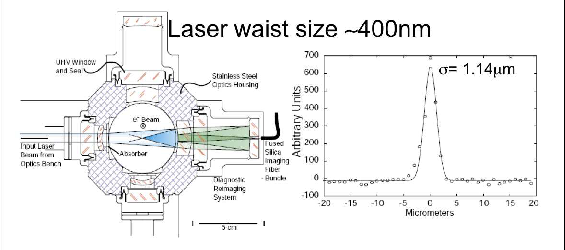} \\
\end{tabular}
\caption{Schematic of a laser wire scanner using an optical cavity around the electron beam instead of a high power laser. The system is installed on a damping ring so the electron beam can be considered practically continuous (left, KEK-ATF). Schematic of a compact laser wire using reflecting optics inside the vacuum chamber to focus the laser and relative measurement (right, SLAC-SLC).}
\label{fig:lwsatf}
\end{figure}

One alternative to powerful lasers are optical cavities with high quality factor (Q) installed across the beam under measurement. The laser needed in this case is much simpler, however, the optics set--up becomes very complicated; \Fref{fig:lwsatf} shows a schematic of such a system.

The optics used to focus the laser is another very delicate part of the system together with the scanning mechanism based on mirrors and precise actuators like piezoelectric actuators. An alternative to the complex and costly scanning systems consists in moving the  particle beam itself, but this requires a very good control of the beam position, sometimes difficult to achieve.

\subsection{Two--dimensional sampling}
\subsubsection{Scintillating screens}
Especially during the commissioning of an accelerator or of a beam line, it is important to have the possibility of acquiring two--dimensional distributions of the particles, or, in other words, it is necessary to take a picture of the beam. As described earlier, 1D profiles can be sufficient to tune and characterize a particle beam if its overall properties are well known and understood. When this is not the case 1D profiles do not give enough information to allow identification of artefacts and problems. Scintillating screens are in fact one of the earliest used diagnostics in particle accelerators. Initially the screens were observed directly by human eyes, when the energy and intensity of the beams became dangerous the direct observation was replaced by TV cameras.

The principle of scintillating screens as described earlier is that charged particles traversing a material ionize and excite the atoms or molecules inside. Part of the energy deposited in the material  in this process is then returned under the form of light. On the screen, at each location, the amount of light emitted is proportional to the number of particles that crossed it. For a monitor to work properly the linearity between particle density and light emission is of the utmost importance.

Under the label of scintillating screens there are in reality a huge number of different devices designed for different uses. In some cases thin coatings of light emitting materials (phosphors) are deposited on rigid substrates, usually for low energy and low intensity beams; the substrate can also be glass offering the opportunity to look from the back side. More often the scintillating plates are rigid objects made of some sort of ceramic, quartz, or monocrystals.

Among the most used ceramics it is worth mentioning aluminium oxide (alumina) and among the monocrystals YAG, a material also used in the fabrication of lasers. Often these materials are doped in order to increase the light emission or shift the wavelength of the emission for a better coupling with the light detector. In the case of alumina the most frequently used dopant is chromium and the material goes under the name of \emph{chromox}. Alumina screens are by far the most used in particle accelerators. This material is in fact very robust both mechanically and thermally, has a very good light emission yield, and is relatively cheap.

\begin{figure}[htb]
\centering\includegraphics[width=0.3\linewidth]{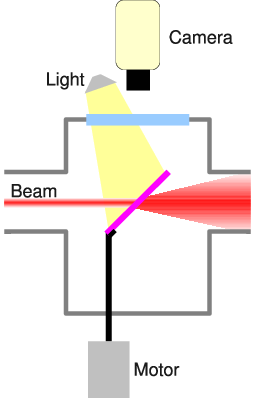}
\caption{Schematic of a scintillating screen monitor}
\label{fig:mtv}
\end{figure}

A profile monitor based on scintillating screens is typically composed of the screen, an insertion mechanism, an illumination system, and a video camera for the detection as can be seen in \Fref{fig:mtv}. As already mentioned it is very important that the relation between particle density and acquired signal be linear. This means that the scintillating screen itself must have a linear curve of emission, but also that the detector must have a linear response and that no saturation occours. It is very frequently the case that the image detector becomes saturated leading to a distorted distribution. For this reason selectable neutral density filters in front of the camera are often used.

Scintillating screens are usually rather thick, of the order of one or more millimetres. The multiple scattering occurring inside the material increases the divergence of the beam and induces beam losses. This is the one of the reasons why the screens are inserted only when required and otherwise retracted.

Often the screen is installed at 45$^\circ$ to the beam and the camera at 90$^\circ$ as depicted in \Fref{fig:mtv}. This solution is convenient because it minimizes the longitudinal space required for the monitor and a round beam will appear round on the image, this is without considering the optical aberrations.
In reality when the camera is not perfectly orthogonal to the screen a trapeze deformation of the image occurs, this is due to the different distances between the screen and the camera at different locations on the screen.

\begin{figure}[htb]
\centering\includegraphics[width=0.7\linewidth]{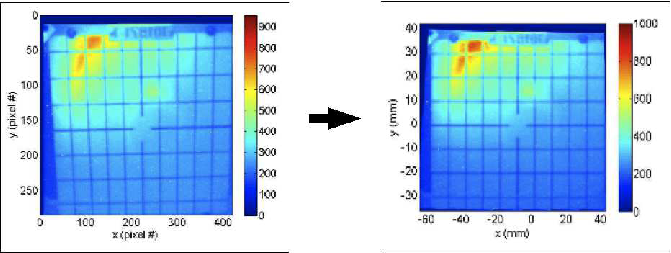}
\caption{Correction of geometrical aberrations due to the optics system and misalignment. Before correction (left) and after (right).}
\label{fig:mtvcorrections}
\end{figure}

Calibration marks on the screen can be very useful, provided they do not interfere with the measurements, as they help in identifying errors and can be used to develop correction algorithms \cite{Benedetto:1102705} (\Fref{fig:mtvcorrections}).

\begin{figure}[htb]
\centering\includegraphics[width=0.5\linewidth]{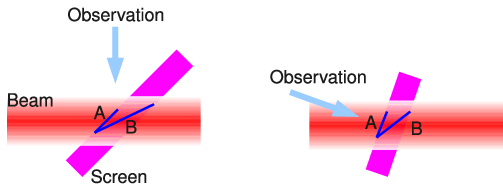}
\caption{Errors due to the finite size of the screen thickness: A indicates the desired observation while B indicates the real observation.}
\label{fig:mtvaberrations}
\end{figure}

Another type of aberration on the image is caused by the finite thickness of the screen. As can be seen in \Fref{fig:mtvaberrations}, the photons are emitted all over the volume traversed by the beam and the angle of observation has an effect on the image observed. This effect is very important for mono crystal screens like YAG where one has to carefully consider the effect of refraction when the light exits the screen \cite{Ischebeck:2015xua}.

The best solution would be to have the screen orthogonal to the beam and the camera orthogonal to the screen, unfortunately this would mean having the camera or at least a mirror placed on the beam path. Observing the back side of the screen either directly (low energy machines) or using a mirror is a technique used in some cases. A more general compromise is to tilt the screen at small angles (\Fref{fig:mtvaberrations} right), the drawback is that the longitudinal occupancy of the monitor on the beam line is increased, not always a possibility. 

\begin{figure}[htb]
\centering\includegraphics[width=0.5\linewidth]{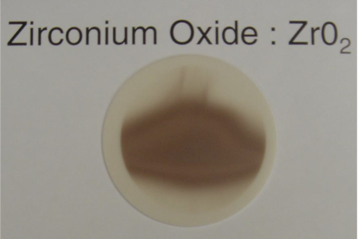}
\caption{Zirconium oxide screen damaged by an intense Lead beam. The discoloration arises from the change of the oxide composition affecting the light emission.}
\label{fig:damagedbtvscreen}
\end{figure}

The energy deposited by the beam in the screen can be sufficient to create a permanent damage. An example of damaged scintillating screen is shown in Fig.~\ref{fig:damagedbtvscreen}.

The instantaneous temperature increase in a material caused by an impinging beam can be expressed as
\begin{equation}
\Delta T(x, y)= \frac{\frac{dE}{dx} \rho\, i(x,y)} {\rho\, c_v}= \frac{\frac{dE}{dx} i(x,y)} {c_v},
\label{eq:temperaturestep}
\end{equation}
where $dE/dx$ is the energy deposition from the Bethe--Bloch formula, $\rho$ is the material density. $i(x,y)$ is the particle density of the impinging beam and $c_v$ is the specific heat of the material of which the screen is made of. It is interesting to note that the only material properties of importance for this effect are the specific heat and the stopping power. In fact the stopping power, in the form given by Bethe--Bloch, i.e. normalised by the material density, is very similar for all materials and depends only on the projectile characteristics. From this we can conclude that, from the thermal point of view, the higher the product $T_{\text{max}} c_v$ the more resistant the material is. In reality one has to consider all the thermo--mechanical effects as temperature-gradient-induced-stresses can break the material well below the fusion or sublimation temperature.

\subsubsection{Optical transition radiation screens}
The scintillating screens described above have two main limitations: the linearity of the response curve, also due to saturation in the material, and the time persistency of the emission. As described already the linearity is an essential condition for profile monitors. The decay time of the emission is usually less important, but there are applications where a short slice of the beam, or a limited number of bunches, have to be acquired. In this case most scintillators would fail as a few microseconds is the fastest decay time available, and even in these fast materials the decay may be composed of many time constants some much longer than the few microseconds of the main emission line. Long decay times are often referred to as persistence and can be many seconds or even minutes.

In a previous section, the \v{C}erenkov radiation was described and although it would solve the linearity and time response limitations of the scintillating screens, it poses more problems than it solves owing to the geometry of the emission.

\begin{figure}[htb]
\centering\includegraphics[width=0.3\linewidth]{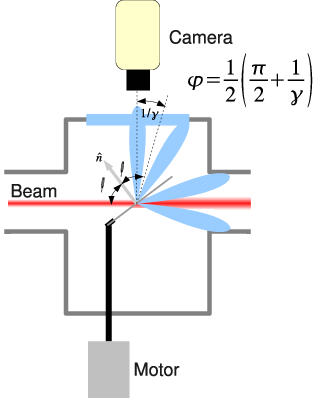}
\caption{Schematics of an OTR monitor, note how the angles between beam, screen and camera have to follow a precise scheme.}
\label{fig:otrschem}
\end{figure}

\begin{figure}[htb]
\centering\includegraphics[width=0.5\linewidth]{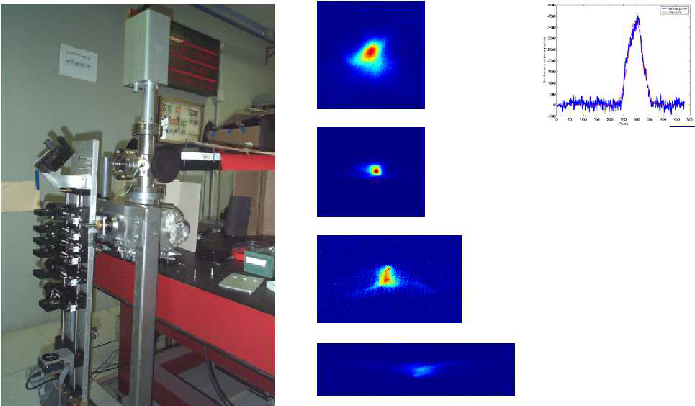}
\caption{Picture of an OTR station used at DESY and images acquired with this type of system on TTF. Note how the complexity of the particle distribution could not be deduced from 1D profiles alone.}
\label{fig:otrttf}
\end{figure}

However, a third type of radiation exists that can be used in place of the scintillating screens: optical transition radiation. OTR, described in a previous paragraph, is a surface emission (contrary to the volume emission of scintillation and \v{C}erenkov), it has a characteristic angular distribution and is radially polarized. \Figure[b]~\ref{fig:otrschem} shows a schematic of an OTR system based on backward emission (forward radiation could in principle also be used, but it is difficult to separate the light from the particles and for this reason it is rarely used). \Figure[b]~\ref{fig:otrttf} shows an OTR station installed in TTF at DESY together with a few beam images acquired with this device.

Installing the camera just on the reflection direction of the beam would mean looking in the hole of the emission (centre of the cone) with almost no light in the aperture of the lens. The solution is to install the camera at a slightly different angle so that it is centred on the peak of the emission (portion of the OTR lobe). The angle of the OTR cone is $1/\gamma$ and thus the correct angle between the camera and the screen and the screen and the beam, these are of course equal because of the mirror--like effect, is given by
\begin{equation}
\varphi=\frac{1}{2}\left( \frac{\pi}{2}\pm\frac{1}{\gamma} \right) .
\label{eq:otrobservation}
\end{equation}

The fact that the emission is not isotropic can introduce artefacts in the image as particles hitting different areas of the radiator or with different angles of incidence can have different acceptances (the fraction of emitted light collected by the camera). When designing an OTR system it is important to study this effect, optical simulation packages are very useful in this regard.

Another important parameter in OTR systems is the reflectivity of the radiator as we use the backward emission. The surface properties of the radiator influence both the total amount of light, with the total reflection coefficient, and the angular distribution, a rough surface will diffuse the light and smooth the peculiar OTR distribution. In some cases, for example when large screens are needed,  the OTR radiator is purposely rough so as to smear the emission and obtain a more uniform angular distribution. Another solution for these applications consists in using radiators with parabolic shapes; if properly designed, parabolic radiators can concentrate the emission toward the camera independently of the position on the screen.

\begin{figure}[htb]
\centering\includegraphics[width=0.6\linewidth]{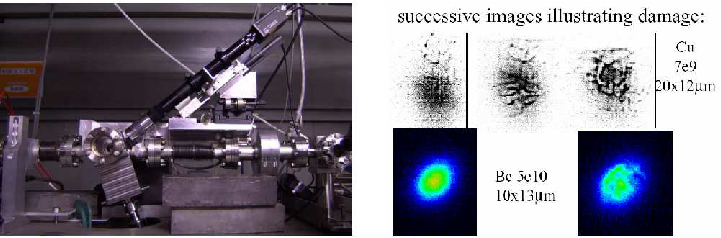}
\caption{Picture of an OTR station used at KEK and images acquired with this type of system on ATF. The images show the damage caused to the radiators (copper and beryllium) by the beam in only a few shots.}
\label{fig:otratf}
\end{figure}

Frequently OTR radiators are made with high--quality mirror surfaces, using either metallic foils or thin substrates coated with metallic depositions. The second solution often offers better mirror qualities while the first reduces the amount of material on the beam path. Typical metallic foils are aluminium or titanium down to a few micrometres thickness while typical metallized substrates are silicon or silicon carbide a few hundred micrometres thick (like the wafers used in microelectronics). Aluminium or gold are frequently used as coating materials. Carbon foils are also used where high density beams have to be observed thanks to the excellent thermal properties of graphite (\Fref{fig:otratf} shows how a high density beam can destroy a radiator), on the other hand the optical qualities are not quite perfect.

\begin{figure}[htb]
\centering\includegraphics[width=0.6\linewidth]{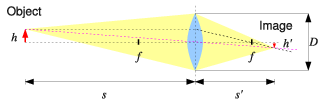}
\caption{Simple optical system composed of a single thin lens}
\label{fig:opticalsystem}
\end{figure}

If the numerical aperture of the optical system is large enough, and this also depends on the $\gamma$ of the beam, the whole OTR cone can be observed. In order to understand this better we shall look at a practical case.
\Figure[b]~\ref{fig:opticalsystem} shows a simple optical system made of a single thin lens. We first define the magnification factor of the optical system based on the size of the beam to be observed and the size of the camera sensor. Then we choose the distance between the object (the screen) and the lens, this is usually limited by the presence of the vacuum tank. At this point we can calculate all the remaining parameters of the optical system.

The magnification factor of the system $m$ is given by
\begin{equation}
m=\frac{h^\prime}{h}=\frac{s^\prime}{s} ,
\label{eq:opticsm}
\end{equation}
and using the lens maker formula
\begin{equation}
\frac{1}{f}=\frac{1}{s}+\frac{1}{s^\prime} ,
\label{eq:opticslens}
\end{equation}
we can deduce
\begin{equation}
f=\frac{sm}{m+1} .
\label{eq:opticsf}
\end{equation}

We can use  \Eref{eq:opticsf} to calculate the focal length for the lens and then \Eref{eq:opticssp} to calculate the last unknown parameter, the distance between the lens and the camera
\begin{equation}
s^\prime=\frac{sf}{s-f} .
\label{eq:opticssp}
\end{equation}

In a practical case we can have $s= 300\Umm$ and $m=0.2$ from which we can calculate $f= 50\Umm$ and $s^\prime= 60\Umm$.
In many cases we shall not use a single lens as this introduces aberrations, but an optical system based on many lenses. In our example we assume the use of a ready--made camera lens. A good CCTV $50\Umm$ lens has $f_{/\#} \sim 1.4$, from this we can calculate the lens diameter (being a complex lens system this is in fact the clear entrance aperture of the lens and not the physical diameter of the first lens)
\begin{equation}
D=\frac{f}{f_{/\#}}=\frac{50}{1.4}=36\Umm ,
\label{eq:opticsD}
\end{equation}
and the acceptance angle at the centre of the screen
\begin{equation}
\theta=\frac{\frac{D}{2}}{s}=\frac{36}{600}=0.06\Urad .
\label{eq:opticsaperture}
\end{equation}
We can now calculate the relativistic $\gamma$ corresponding to this aperture for the OTR radiation:
\begin{equation}
\gamma_{min}=\frac{1}{\theta}= \frac{1}{0.06}=17 .
\label{eq:opticsgammamin}
\end{equation}

From \Eref{eq:opticsgammamin} we obtain that for the setup of the example and for beams with $\gamma$ larger than $17$ we can observe the whole OTR cone, i.e., we can centre the camera on the reflection axis of the OTR radiator, while for beams of lower momentum we have to centre the camera on one of the lobes as described before.

\subsubsection{Synchrotron light monitors}
Another way of sampling the two--dimensional distribution of the particles is offered by synchrotron radiation (SR). The advantage of synchrotron radiation over OTR is that it does not require a radiator. This means that there are no limitations on the particle densities that can be observed. Moreover it also allows the continuous observation of the beam profiles.

SR monitors are almost always designed around existing magnets, that is around magnets that are already installed in the machine for other purposes. It may be necessary, however, to install dedicated magnets in order to overcome specific needs. For example, if the $\gamma$ of the particle is too low, the spectrum of the emission from a lattice bending magnet can be concentrated in the infra-red and longer wave lengths, with important difficulties for detection. In a case like this a special magnet, called an undulator, can be used.

Undulators are periodic magnets in which the central wavelength of the SR emission depends on the magnetic period. Undulators can also be used if the intensity of the emission is not sufficient as they can be much more effective than a simple bending magnet in generating SR. Of course different beam conditions will require different solutions. It is important to keep in mind that even if an undulator is used as source a dipole magnet is still required to separate the light from the particles. Sometimes the edge effect in bending magnets can be used to shift the SR spectrum. The SR emission at the entrance and exit edges of the magnet is in fact slightly shifted towards shorter wavelengths since SR pulses are shorter there. In some cases this shift can be sufficient.

One of the biggest problems in building a SR profile monitor is the fact that the source is not well defined. As the light is emitted in the interaction between the particles and the magnetic field, the particles will emit during the whole time they traverse the magnetic field region, this can be almost continuous in compact circular machines where bending magnets are very close to one another. The result is like taking the picture of a moving car at night on a bend, we see the trace of the car, but can not make out the details of the headlights. In the case of the car it is possible to use a very fast shutter time to compensate for the motion, in a SR monitor this is not possible as the light and the particles travel almost at the same speed.

\begin{figure}[htb]
\centering\includegraphics[width=0.6\linewidth]{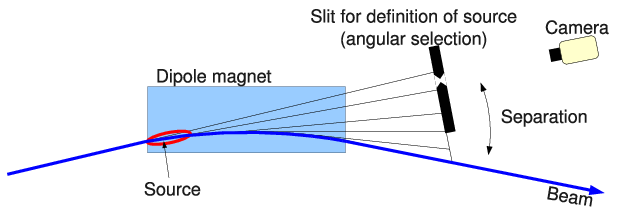}
\caption{Sketch of a synchrotron light monitor using a bending magnet as source. The source is delimited using an angular selection defined by the slit and the natural small opening angle of SR.}
\label{fig:synclite}
\end{figure}

Another possibility is to use an angular selection and detect only photons that come from a particular point, our source. To define an angle we usually need two slits, only photons with the right angle will pass through both. In a SR monitor we can use the fact that the light is emitted in a very narrow cone in the direction of motion of the particle and reduce the system to a single slit \Fref{fig:synclite}. 

Another important problem of SR is the diffraction. From optics we know that there is a precise relation between the resolution of an optical system and the numerical aperture. The finite aperture introduces diffraction and the influence of the diffraction on the image increases as we reduce this aperture.

For an optical system the diffraction limited spot size (image plane) is often defined as the diameter of the Airy disk
\begin{gather}
\sin{\theta}= 1.22 \frac{\lambda}{D} \label{eq:airyangle} ,\\
a= \sin{\theta}\, s^\prime= 1.22 \frac{\lambda}{D} s'= 1.22 \frac{\lambda}{D} m\, s\label{eq:airydisk} ,
\end{gather}
with $\theta$ the angular resolution, $a$ the diameter of the Airy disk, $\lambda$ the observation wavelength, $D$ the clear aperture of the lens, $s$ the distance between the source and the lens, $s^\prime$ the distance between the lens and the image and $m$ the magnification of the optical system. For our purposes it is useful to approximate the radial distribution of the Airy diffraction pattern with a Gaussian, the best fit is given by
\begin{gather}
I(x)\approx I_0 e^{-\frac{x^2}{2\sigma_d^2}} \label{eq:diffractiongauss}, \\
\sigma_d \approx \frac{a}{2.9} = 0.42 \frac{\lambda}{D} m\label{eq:diffractionsigma},
\end{gather}
and transposing the result to the object side, dividing by the magnification, we obtain
\begin{equation}
\sigma_{\text{d-object}}=0.42 \frac{\lambda}{D} \label{eq:diffractionsigmaobject}.
\end{equation}

Usually in synchrotron radiation monitors the clear aperture of the optical system is larger than the aperture of the radiation itself, that is all the light emitted at a given source point is collected. In this case the diffraction limit is defined by the aperture of the radiation instead of the clear aperture of the system. We can approximate the SR angular distribution with a Gaussian of width $\sigma_\theta$, in this case the fit to the diffraction pattern will give
\begin{equation}
\sigma_{\text{d-object}} \approx 0.15 \frac{\lambda}{\sigma_\theta},
\label{eq:diffractionsigmaobjecttheta}
\end{equation}
and applying \Eref{eq:synchopeningangle} we obtain
\begin{equation}
\sigma_{\text{d-object}} \approx 0.19 \frac{\lambda \gamma}{C(y)}.
\label{eq:diffractionfinal}
\end{equation}

The important result of \Eref{eq:diffractionfinal} is that the resolution of the synchrotron light monitor is proportional to $\lambda\gamma$, this means that for highly relativistic beams we can expect serious resolution limits that can only be solved by observing shorter and shorter wavelengths. For this reason, on electron machines, where small resolutions are required, even below the micro meter, synchrotron light monitors are often based on the detection of soft X-rays. Other techniques like deconvolution of the point spread function or interferometry allow one to extract useful information even when the diffraction effects are important.

Another important aspect of \Eref{eq:diffractionfinal} is that it is independent of the optical parameters of the system. This means that the optical resolution that can be achieved with a synchrotron light monitor is defined by the radiation source alone. In reality diffraction is not the only effect contributing to the optical resolution, other effects like the depth of field are also important. The design of the optics still requires careful studies and simulations.

Let us examine a practical example, a synchrotron light monitor for a typical SR light source with $E=2\UGeV$ ($\gamma= 3914$) and $\rho= 10\Unit{}{m}$. The critical energy of the emitted photons is $E_{cr}= 1.7\UkeV$ and for $\lambda= 400\Unm$ we obtain $C(y)\approx 9$.
Plugging these values into \Eref{eq:diffractionfinal} we obtain a resolution of $\sigma_d \approx 33\Uum$. It is important to note that when the observation wavelength is far from the critical wavelength the $C(y)$ parameter can have a large impact on the estimation of the resolution and can not be neglected.

In SR monitors one critical component is the mirror used to extract the light out of the vacuum chamber. In lepton machines the SR spectrum can be very hard, with $\lambda_{cr}$ in the hard X-rays, and very intense. This leads to a large energy deposition inside the mirror creating gradients that distort the shape and even potentially damaging the surface. Often this problem is mitigated by placing an absorbing mask on the central path of the radiation, where the hard components are predominant, and using only the softer and less intense outer region for the detection. If the extraction mirror has to be placed near the circulating beam the radio frequency fields generated by the circulating particles can heat the mirror or induce sparking, similar to the disasters that can be obtained by placing metallic objects in a microwave oven. In this case the choice of the geometry and of the materials is very important. 

\begin{figure}[htb]
\centering\includegraphics[width=0.6\linewidth]{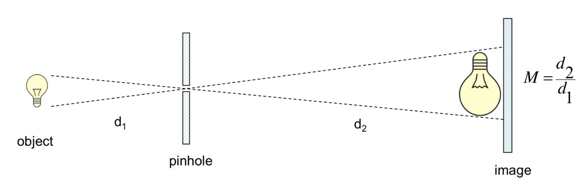}
\caption{Schematic of the camera obscura, the light emitted by the source passes trough a small hole and is geometrically projected onto a screen.}
\label{fig:pinholecamera}
\end{figure}

Synchrotron radiation monitors for the X-rays are complicated by the difficulty of deflecting and focusing the radiation in an imaging system. X-rays mirrors only work at grazing angles and precise wavelength filters are very complicated (Bragg monochromators). Four main techniques are available: pinhole camera, X-rays refractive lenses, Fresnel plates and focusing mirrors. All these techniques have pro and cons, each one suiting best a particular problem. The most used one is the pinhole camera as it is the simpler to implement. The principle is the same used at the beginning of photography, known as camera obscura, and is described in Fig.~\ref{fig:pinholecamera}. The magnification of the system is given by the ration between $d_2$ and $d_1$ and the depth of field is infinite as there is no focusing element, so the screen can be put at any distance.

The resolution of the pinhole camera is limited by the finite size of the pinhole (blur) and by the diffraction of the pinhole. Assuming we can describe both effects as Gaussian blurring with width $\sigma_{\text{blur}}$ and $\sigma_{\text{diffraction}}$ in case of a Gaussian source we can write \cite{Elleaume:1995jsr}
\begin{equation}
    \sigma_{\text{image}}^2= \sigma_{\text{source}}^2 m^2 + \sigma_{\text{blur}}^2 + \sigma_{\text{diffraction}}^2 \label{eq:pinholespotsize},
\end{equation}
with
\begin{align}
    \sigma_{\text{blur}}&= \frac{w}{\sqrt{12}} \frac{d_1+d_2}{d_1} \label{eq:sigmablur},\\
    \sigma_{\text{diffraction}}&= \frac{\sqrt{12}}{4 \pi} \frac{\lambda d_2}{w} \label{eq:sigmadiff},
\end{align}
where $m$ is the magnification $d_2/d_1$, $w$ the width of the rectangular pinhole, $\lambda$ the wavelength, $d_1$ the distance source--pinhole and $d_2$ the distance pinhole--screen.
The pinhole camera does not require monochromatic light like the X-ray refracting optics, a simple metal plate can be used to remove the low energy photons that would increase the diffraction blur.

\begin{figure}[htb]
\centering\includegraphics[width=0.6\linewidth]{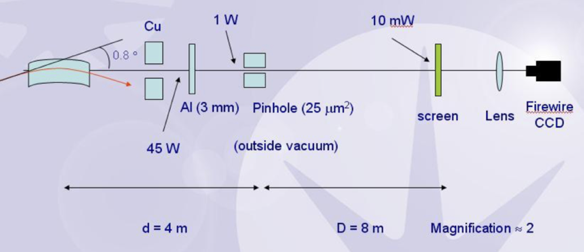}
\caption{Schematic of the X-rays pinhole camera developed by the DIAMOND light source in the UK.}
\label{fig:pinholecameradiamond}
\end{figure}

Figure~\ref{fig:pinholecameradiamond} shows the schematics of a real X-rays pinhole camera developed for a synchrotron light source. As it can be seen it is quite more complicated than the original camera obscura.

Another often used technique to overcome the limitations imposed by diffraction is to use a double slit interferometer like the one used by Young. The visibility of the interference fringes (interferogram) can be used to calculate the spatial coherence of the source and infer its transverse dimension.

\section{Light sensors}
Many of the monitors described above are based on the detection of visible light. Light sensors can be divided into two families:
\begin{itemize}
\item One--dimensional sensors
\begin{itemize}
\item Photodiode array
\item Linear CCD
\item Segmented photomultiplier (multi--anode PMT)
\end{itemize}
\item Two-dimensional sensors
\begin{itemize}
\item Area CCD
\item Area C-MOS
\item Multi anode photomultipliers
\end{itemize}
\end{itemize}

\subsection{One--dimensional sensors}
One--dimensional sensors are usually faster. The readout can be of sequential or parallel type. In the second case each channel is read out independently and the acquisition chain requires as many channels as there are channels in the detector. The sequential type instead requires only one readout channel and the values of the channels of the detector are acquired one after the other using a multiplexing technique.

Usually photodiode arrays and multi--anode PMTs are of the parallel type while the linear CCDs are of the sequential type. The parallel type can only be used when the number of channels is limited as the electronics and cabling would otherwise become too complex, typically up to 32 channels. The advantage of these devices is that they can be very fast, with acquisition rates up to 100\UMHz{}. The linear CCDs on the other hand allow a large number of channels with a rather simple readout electronics.

CCDs with over 1000 channels are available, the readout speed is, however, much reduced and usually does not exceed 100\UkHz{} depending on the number of channels (typical channel data rate is of the order of 20\UMHz{}). Often linear CCDs allow the selection of the number of channels to be read depending on the needs of the application. It is then possible to develop a detector that can switch between fast acquisition with few channels and slow acquisition with many channels.

In terms of sensitivity the multi--anode PMTs are superior to the other two devices as an amplification up to $10^5$ takes place inside the device itself allowing the detection of single photons. The diode arrays and the linear CCDs have similar sensitivities and require a large number of photons for a good signal ($\sim$50 000/channel).

The PMTs on the other hand are a bit more complicated to use as they require a high voltage bias ($\sim$1000\UV{}) and can be damaged by a high light intensity. PMTs are also very good in terms of radiation resistance, much better than the other two solid--state devices. 

\subsection{Two--dimensional sensors}
Two--dimensional detectors only have sequential readout as the minimum number of channels needed to acquire an image is already much too large for parallel readout ($32\times 32= 1024$). Although 2D segmented photomultipliers and photodiode arrays exist, their granularity is usually very small and not made for the acquisition of images (max. $8\times 8$).

There are two main families of solid--state 2D image sensors, CCDs and C-MOS. The difference between the two is that the CCD is a matrix of pixels (electron wells) that are shifted one after the other towards the readout electronics while the C-MOS is a matrix of pixels each one directly connected via a multiplexing system to the readout electronics. In principle the difference should not be important for normal acquisitions, in fact the CCD is just a silicon substrate with \emph{transparent} electrodes on top, while the C-MOS device requires electronic components next to each pixel. The filling factor of the two devices is thus quite different (fraction of the area of the detector that detects the impinging photons). Innovation in microelectronics in the last 10 years and clever designs have improved dramatically the performance of the C-MOS sensors. C-MOS cameras, once cheap low end devices, have now overtaken the CCD in almost all application fields. In the near future the CCD technology will completely disappear or be relegated to very special market niches. State of the art commercial C-MOS sensors have achieved read-out noise of few electrons, almost a factor ten better than equivalent CCDs. C-MOS sensors optimised for low light conditions, called scientific C-MOS or sC-MOS, can achieve readout noise of less than one electron, the cost is however one order of magnitude larger than standard commercial sensors.

One advantage of the C-MOS is that any portion of the detector can be read out at will while on CCDs the whole matrix has to be read out. This is a big advantage for us as we can readout an area of interest (AOI) at much higher rate. Readout speeds up to the kilo Hertz are possible with simple cameras for an AOI of $\approx 100\times 100$ pixels.

Readout speeds of 2D devices are quite slow, typical sensors are designed for 50\UHz{} readout. There are, however, special C-MOS devices designed for higher acquisition speeds. In these devices the readout is split into many parallel blocks (called taps) to reduce the time needed, the use of small AOI can further increase the frame rate. With such special cameras speeds of the order of 1\UMHz ($10^6$ frames per second) are possible.

A interesting 2D device is the video tube. Nowadays this technology is obsolete and it is quite difficult to find new parts; it is, however, still widely used in particle accelerators as it is very tolerant of ionising radiation. In many locations inside accelerators the radiation is too high for devices like CCDs or C-MOS and the video tube is the only device that can acquire images of the beam in these situations.

There is also another type of image sensor, similar to the C-MOS, the CID (charge injection device). Although this technology has not been pushed as far in performance as the C-MOS it has the advantage that it has been optimised for radiation--hardness. These devices can be used in areas were radiation is moderated, usually rad--hard CID cameras can stand up to 5\Urad[M]{} while CCDs and C-MOS only survive up to $\sim$100 \Urad[k]{}.

There are a number of projects ongoing world wide that aim at producing solid state radiation hard image sensors. The hope is to obtain a replacement for the video tubes with better performances.

Another important aspect of image cameras is the format of the output signal. Until few years ago the standard was a well defined analog signal that was then acquired by mean of so called frame grabber cards. With the ascent of C-MOS sensors the digitisation has been embedded in the sensor chip itself so that the cameras already output a digital signal typically over a USB or ethernet interface.

\section*{Acknowledgements}
I would like to thank all the people working in beam instrumentation as most of the material presented in this lecture has been taken from many published documents. It is impossible to mention specific names as there are so many involved.

\section*{}

\end{document}